\documentclass[fleqn,usenatbib]{mnras}

\usepackage{newtxtext,newtxmath}

\usepackage[T1]{fontenc}

\DeclareRobustCommand{\VAN}[3]{#2}
\let\VANthebibliography\thebibliography
\def\thebibliography{\DeclareRobustCommand{\VAN}[3]{##3}\VANthebibliography}

\usepackage{graphicx}	
\usepackage{amsmath}	
\usepackage{newtxtext,newtxmath}
\usepackage{relsize}
\usepackage{wrapfig, subcaption, booktabs, longtable, caption, makecell}
\usepackage{tikz} 

\newcommand{\cosmounits}{\,$\rm{mK^2}\,\it{h^{-3}} \,\rm{Mpc^{3}}$}	

\title[SNRs, Centaurus A and the 21\,$\rm{cm}$ EoR Signal]{Investigating the Contribution of Extended Radio Sources to the Epoch of Reionisation Power Spectrum}

\author[J.~H. Cook et al.]{
Jaiden H. Cook,$^{1,2}$\thanks{E-mail: Jaiden.cook@curtin.edu.au}
C. M. Trott,$^{1,2}$
J. L. B. Line$^{1,2}$
\\
$^{1}$International Centre for Radio Astronomy Research, Curtin University, Perth, Australia\\
$^{2}$ARC Centre of Excellence for All Sky Astrophysics in 3D (ASTRO 3D)\\
}

\date{Accepted XXX. Received YYY; in original form ZZZ}

\pubyear{2021}

\begin{document}
\label{firstpage}
\pagerange{\pageref{firstpage}--\pageref{lastpage}}
\maketitle

\usetikzlibrary{decorations.pathreplacing}
\usetikzlibrary{calc,patterns,angles,quotes}

\begin{abstract}
We investigate the contribution of extended radio sources such as Centaurus A, and Galactic supernova remnants (SNRs) to our ability to detect the statistical $21\,\rm{cm}$ signal from the Epoch of Reionisation (EoR) with the Murchison Wideﬁeld Array (MWA). These sources are typically ignored because they are in highly attenuated parts of the MWA primary beam, however in aggregate these sources have apparent flux densities of $10\,\rm{Jy}$ on angular scales we expect to detect the $21\,\rm{cm}$ signal. We create bespoke multi-component 2D Gaussian models for Galactic SNRs and for Centaurus A, and simulate the visibilities for two MWA snapshot observations. We grid those visibilities and then Fourier transform them with respect to frequency, averaging them both spherically and cylindrically to produce the 1D and 2D power spectra. We compare the simulated 1D power spectra to the expected 21\,$\rm{cm}$ power spectrum. We find that although these extended sources are in highly attenuated parts of the MWA primary beam pattern, collectively they have enough power ($\sim10^4-10^5$\cosmounits) on EoR significant modes $(|\mathbf{k}| \lesssim 0.1\,h\,\rm{Mpc^{-1}})$ to prohibit detection of the 21\,$\rm{cm}$ signal ($\sim10^4$\cosmounits). We find that $50-90\%$ of sources must be removed in order to reduce leakage to a level of $\sim10-20\%$ of the 21\,$\rm{cm}$ power spectrum on EoR significant modes. The effects of widefield extended sources will have implications on the detectability of the 21\,$\rm{cm}$ signal for the MWA and with the future Square Kilometre Array (SKA).
\end{abstract}


\begin{keywords}
cosmology: dark ages, reionization, first stars -- techniques: interferometric -- methods: statistical -- stars: supernovae: general
\end{keywords}




\section{Introduction}

Radio observations of the redshifted 21\,$\rm{cm}$ neutral hydrogen emission line have the capability to reveal underlying astrophysical formation mechanisms during the cosmic dawn, and the Epoch of Reionisation (EoR) \citep{Furlanetto_2006}. The EoR is the period of cosmic time where the predominantly neutral hydrogen inter-galactic medium (IGM), transitioned to a fully ionised state after the formation of the first stars, galaxies, and black holes. Observations of quasars \citep{Fan_2006} and the anisotropies in the Cosmic Microwave Background through the Sunyaev–Zel’dovich effect \citep{SZ-effect}, have constrained the EoR to a redshift range of $5.4 \lesssim z \lesssim 10$. The cosmological nature of the 21\,cm emission line allows for the direct observation of the full reionisation history. The future Square Kilometre Array (SKA) promises to directly image the redshifted 21\,$\rm{cm}$ signal during the EoR \citep{SKA-EoR}.

The current generation of low frequency radio instruments lack the sensitivity to directly image the 21\,$\rm{cm}$ signal, and are thus focused on estimating the 21\,$\rm{cm}$ statistics as a function of spatial scale by calculating the 21\,$\rm{cm}$ power spectrum. The 21\,$\rm{cm}$ statistics have the potential to differentiate between different reionisation scenarios, and therefore provide an insight into the underlying astrophysical reionisation mechanisms \citep[see][for comprehensive reviews]{Furlanetto_2006,Morales_rev_2010,Pritchard_2012,Furlanetto2016}. The current generation of radio instruments includes the Murchison Widefield Array \citep[MWA,][]{MWA-sci,MWA-PhaseI,MWA-PH2}; Low-Frequency Array \citep[LOFAR,][]{LOFAR}; the Precision Array for Probing the Epoch of Reionization \citep[PAPER,][]{PAPER}; Hydrogen Epoch of Reionization Array \citep[HERA,][]{HERA}; The Amsterdam–ASTRON Radio Transients Facility and Analysis Center \citep[AARTFAARC,][]{AARTFARRC}; the New extension in Nancay upgrading LOFAR \citep[NenuFAR,][]{NenuFAR}. The MWA is a 256 element interferometer, with 128 operational at any one time in a compact or extended configuration \citep{MWA-PH2}. Measuring the statistical 21\,$\rm{cm}$ signal from the EoR is one of the main science goals of the MWA \citep{MWA-sci}.

Foreground Galactic and extra-Galactic radio sources at redshifted 21\,$\rm{cm}$ frequencies pose a fundamental problem for detecting the 21\,$\rm{cm}$ signal during the EoR. These foreground sources can be $10^4-10^5$ times brighter than the underlying cosmological 21\,$\rm{cm}$ signal \citep{Furlanetto_2006}. The frequency structure of the 21\,$\rm{cm}$ signal varies rapidly with frequency when compared to foreground emission \citep{Shaver_1999}. Foreground emission is proportional to a power law distribution, and varies relatively smoothly over frequency. Therefore foreground power is expected to be primarily isolated to low line of sight $k$ Fourier modes compared to the 21\,$\rm{cm}$ EoR signal \citep{Morales2004,Bowman2009}. However instrumental chromaticity imparts highly varying spectral structure which leaks power into prospective EoR modes through a process known as mode mixing \citep{Bowman2009,Datta2010}. One way to avoid some of these effects is through the 2D power spectrum, which separates the power spectrum modes into line of sight modes $k_{||}$ and perpendicular angular modes $k_\perp$ in units of $\textrm{Mpc}^{-1}$ \citep{Morales_2006,Datta2010}. Radio interferometers sparsely sample the $uv$ plane (which is proportionate to $k_\perp$), however baseline length is wavelength dependent and so introduces frequency structure into the foreground emission. As a result of this structure, foreground emission leaks into higher $k_{||}$ modes as a function of $k_\perp$ \citep{Morales_2012,Trott_2012,Vedantham_2012}, resulting in a wedge-shaped foreground-dominated area.

Most of the foreground power is contained in the wedge, leaving a relatively clean `EoR window' \citep{Vedantham_2012}. However, calibration errors and primary beam chromaticity can cause leakage from the foreground wedge into the EoR window \citep{Morales_2012,Trott_2012,Barry_2016}. This problem is compounded for sources further from the centre of the field, as the primary beam changes more with frequency the further away from the point of maximum sensitivity. \citet{Foregrounds-1} analysed the effects of including source subtraction from the sidelobes of the MWA primary beam when calculating the 2D power spectrum. They found that sources further from the centre of the field leaked more power from the foreground wedge into the window. The MWA primary beam spectral structure for different EoR fields is shown in Figures $27$ and $28$ in \citet{EoR-limits_Trott2020}. At the edges of the sidelobes, and away from the main lobe, the MWA primary beam spectral index is steep, introducing rapidly changing spectral structure to sources in these locations. Furthermore, \citet{Foregrounds-1} found that including these extra-galactic sources located in the beam sidelobes during foreground removal reduced the power in the EoR window by a few percent.

\citet{Foregrounds-1} was only concerned with point sources in the sidelobes, however, in the EoR\,2\footnote{EoR\,2 field coordinates: RA=$10.3\,\rm{h}$, Dec=-$10^\circ$} field there are several exceptionally bright extended sources, which due to their low apparent surface brightness are generally not included in MWA EoR processing pipelines. Primarily this field contains Centaurus A (CenA), which is the brightest radio galaxy in the sky spanning $4\times8\deg$ with a brightness of $\sim4000\,\textrm{Jy}$ at $183\,\textrm{MHz}$ \citep{Alvarez2000,Ben2013}. CenA is often present or at the edge of one of the MWA primary beam sidelobes for EoR\,2 field pointings. As a result CenA is highly attenuated, but has an apparent brightness on the order of $10\,\textrm{Jy}$. Additionally, the complex spectral structure of the MWA primary beam at the sidelobes imprints frequency structure that can lead to leakage in the EoR window. Leakage at this apparent brightness can still be orders of magnitude brighter than the expected 21\,$\rm{cm}$ signal.

In addition to CenA the Galactic plane also appears in one or several of the MWA primary beam sidelobes. The Galactic plane is populated by a large number of bright supernova remnants (SNRs) as well as large scale diffuse radio emission. SNRs themselves have flux densities that range from $1-1000\,\textrm{Jy}$, and have angular extents that are similar in scale to the expected 21\,$\rm{cm}$ reionisation bubbles \citep{Bubbles_1,Bubbles_2}. Likewise, these sources are in complex parts of the MWA primary beam, which can cause leakage from the foreground wedge into the EoR window. Further complications occur as these extended sources rotate through the MWA primary beam, imparting varying spectral structure in the process. Their extended nature also means the spectral structure imparted by the beam changes across the source, and can vary significantly depending on the location of the source within the primary beam.

The effect of these attenuated but complex sources at the field edge has not been established for 21\,$\rm{cm}$ EoR science. To investigate the amount of leakage caused by these sources in the EoR window, in this work we create a sky-model which contains morphological models of CenA and Galactic plane SNRs. The modelling of the morphological models for Galactic SNRs and CenA is described in Section \ref{sec:data+models}. We then run various sky-models through a simulation pipeline (described in Section \ref{sec:methodology}) which calculates the 1D and 2D power spectrum with a fiducial 21\,$\rm{cm}$ signal \citep[via][]{21cm-FAST}. We then look at how much of the sky-model needs to be subtracted to retrieve the 21\,$\rm{cm}$ signal (Section \ref{sec:results}). In this work we perform all cosmological calculations with the \citet{Plank2018} cosmology, where $h = H_0/100\,\rm{km\,s^{-1}\,Mpc^{-1}}$.

\section{Methodology}\label{sec:methodology}

To test the leakage of Galactic Plane SNRs and CenA into the EoR window, we developed a method which simulates the contribution of extended radio sources to the visibilities measured by the MWA. Briefly, we describe the steps of the method here, going into more detail in the subsequent subsections. The first step generates a sky-model image cube $I(\mathbf{l},\nu)$ as a function of frequency. These sky-model cubes are constructed from multi-component 2D Gaussian models of CenA and Galactic plane SNRs; for details on how the sky-model cubes and the 2D Gaussian model components were created, refer to Section \ref{sec:data+models}. The sky-model cube is Fast Fourier Transformed (FFT) into the Fourier sky-cube $\tilde{I}(\mathbf{u},\nu)$. The visibilities $\mathcal{V}(\mathbf{u},\nu)$ are simulated by sampling the Fourier sky-cube using the MWA $(u,v)$ distribution. The sampling process incorporates the FFT of the MWA primary beam, effectively simulating MWA measurements. The sampled visibilities are then gridded onto the $uv$-plane reconstructing the Fourier sky-cube which is denoted by $\Tilde{\mathcal{I}}(\mathbf{u},\nu)$. An FFT is then performed with respect to the frequency axis to retrieve the reconstructed Fourier sky-cube $\tilde{\mathcal{I}}(\mathbf{u},\eta)$ as a function of the line of sight mode $\eta$. $\tilde{\mathcal{I}}(\mathbf{u},\eta)$ is then averaged both spherically and cylindrically to calculate the 1D and 2D power spectra respectively. 

For comparison a fiducial simulated 21\,$\rm{cm}$ signal is added to a noise only reconstructed Fourier sky-cube $\tilde{\mathcal{I}}_\mathcal{N}(\mathbf{u},\eta)$. This is then spherically and cylindrically averaged to calculate the 1D and 2D noise plus 21\,$\rm{cm}$ signal 1D and 2D power spectra. We then compare the 21\,$\rm{cm}$ signal power spectra to the simulated widefield extended power spectra to determine the significance of leakage at EoR $k$-modes of interest. The fiducial 21\,$\rm{cm}$ signal was generated using \textsc{21cmfast} power spectrum simulations, and is taken from \citet{21cm-FAST}. 

\begin{figure}
    \centering
    \includegraphics[width=0.475\textwidth]{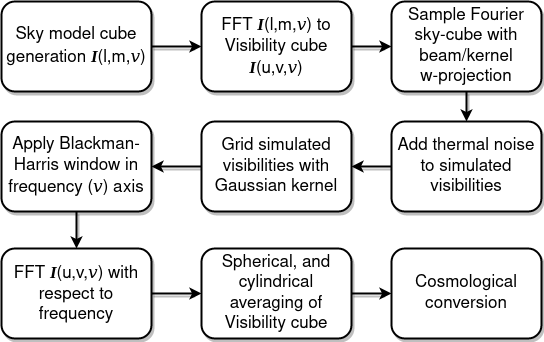}
    \caption{Simulation pipeline flow chart. Shows all the different steps from the sky-model generation to the power spectrum calculation and cosmological conversion.}
    \label{fig:flow-chart}
\end{figure}

To simulate MWA observations we created a simulation pipeline called Observational Supernova-remnant Instrumental Reionisation Investigative Simulator (\href{https://github.com/JaidenCook/OSIRIS}{OSIRIS})\footnote{\href{https://github.com/JaidenCook/OSIRIS}{https://github.com/JaidenCook/OSIRIS}}. The core interferometric simulation functions are based on the \textsc{majick} software package \citep{MAJICK}. The general process of the OSIRIS pipeline is described by the flow chart in Figure \ref{fig:flow-chart}.

\subsection{Fourier Sky Cube}
Radio interferometers measure a complex coherence term known as the visibility $\mathcal{V}(\mathbf{u})$, which is the cross-correlation between two antenna elements. The visibility is described by the measurement equation, which relates the sky-brightness distribution $I(\mathbf{l})$ to the visibility as a function of $\mathbf{u}$ \citep{Radio-synthesis}:

\begin{equation}\label{eq:Vis}
   \mathcal{V}(\mathbf{u},w) = \int^\infty_{-\infty}\frac{B(\mathbf{l})I(\mathbf{l})}{n}\, e^{ -2\pi i\left(w\left(n -1\right)\right)} e^{-2\pi i(\mathbf{u}\cdot\mathbf{l})} d\mathbf{l}
\end{equation}
$B(\mathbf{l})$ is the primary beam as projected onto the celestial sphere, and $n$ is the direction cosine along the phase centre, which is defined by $n=\sqrt{1 - |\mathbf{l}|^2}$. The vector $\mathbf{u}$ represents the physical displacement of the tiles on a plane in units of wavelengths and is represented by the coordinates $(u,v)$; the vector $\mathbf{l}$ contains the direction cosines $(l,m)$ which are defined on the image plane. The $w$-terms encapsulate the curvature of the sky, and are significant because the MWA has a large field of view ($>10\deg$). OSIRIS takes an input sky-model cube $I(\mathbf{l},\nu)$, which is then Fourier transformed with respect to $\mathbf{l}$ via an FFT. The resulting FFT produces the Fourier sky cube $\tilde{I}(\mathbf{u},\nu)$, mapped to a regular $(u,v)$ grid, defined by the extent of the input image $(l,m)$ grid.

\subsection{Simulating Visibilities}\label{sec:degridding}

Simulating the instrumentally-measured visibilites is performed by sampling the $uv$-plane with a kernel that incorporates the MWA primary beam response $B(\mathbf{l},\nu)$ and the curvature of the sky through the $w$-kernel\footnote{The $w$-kernel is defined as $e^{-2\pi i\left(w\left(n -1\right)\right) }$.}. This process samples the Fourier space for each baseline as a function of frequency. The baseline coordinates $(u,v,w)$ for each frequency slice are determined using the MWA Phase I array layout, and a set of \textsc{majick} functions. These functions use the array (east, north, height) and pointing centre to determine the baseline coordinates in meters $(u,v,w)$, which are converted into units of wavelength for each frequency channel. In this work we use a zenith pointed array, since we consider extended radio models of the entire sky. The sampling kernel $\Tilde{K}(\mathbf{u},w_i,\nu)$ for a given baseline at a particular frequency, is the convolution of the FFT of the MWA primary beam, and the FFT of the $w$-kernel:

\begin{equation}\label{eq:kernel-degrid}
    \Tilde{K}(\mathbf{u},w,\nu) = \int^\infty_{-\infty} B(\mathbf{l},\nu) e^{-2\pi i\left(w\left(n -1\right)\right) } e^{-2\pi i(\mathbf{u}\cdot\mathbf{l}) } d\mathbf{l}.
\end{equation}

%
%
The curvature term $w$ is lost in the 2D FFT from image space $(l,m)$ to Fourier space $(u,v)$. The $w$-kernel reincorporates the curvature of the sky through a process called w-projection (see \citet{W-Proj} for further details). Each baseline has a different $w$-term, and as such there is a unique sampling kernel for every baseline. 

%
%
The MWA primary beam $B(\mathbf{l},\nu)$ is generated using the Fully Element Embedded (FEE) model described by \citet{BEAM2016}. The FEE beam model only has a frequency resolution of $1.28\,\textrm{MHz}$, however the channel resolution of the Fourier sky-cube is $\Delta\nu_f = 80\,\rm{kHz}$, thus the FEE beam model requires interpolation as a function of frequency. Without interpolation, the coarse band structure of the beam will be present in the EoR window when we Fourier transform with respect to frequency. Before the OSIRIS pipeline performs the interpolation, the FEE beam model is generated for $36$ coarse channels (bandwidth $1.28\,\textrm{MHz}$) spanning the frequency range $147.2-193.3\,\rm{MHz}$. The resulting beam cube is interpolated as a function of frequency using cubic splines. The observations simulated in this work have a bandwidth of $\Delta\nu = 15.36\,\rm{MHz}$, therefore each simulated observation has $192$ channels. A primary beam model is generated for each channel using the interpolated FEE beam model. 

%
%
%
Using the frequency interpolated FEE beam model, and the $w_i$ term for the $i$th baseline, the OSIRIS pipeline generates a unique sampling kernel for that baseline. The simulated MWA visibility for the $i$th baseline $\mathbf{u}_i$, is determined by taking the sampling kernel weighted average of the $I(\mathbf{u}_j,\nu)$ grid pixels (the subscript $j$ denotes the pixel index) centred at $\mathbf{u}^\prime_i$:

\begin{equation}\label{eq:vis-degrid}
    \mathcal{V}(\textbf{u}_i,w_i,\nu) = \frac{\mathlarger{\sum}\limits_{j=0}^{N}\Tilde{K}(\textbf{u}_j - \mathbf{u}^\prime_i,w_i,\nu) \tilde{I}(\mathbf{u}_j,\nu)}{\mathlarger{\sum}\limits_{j=0}^{N} \Tilde{K}(\textbf{u}_j - \mathbf{u}^\prime_i,w_i,\nu)} .
\end{equation}

The sampling kernel $\Tilde{K}(\textbf{u}_j - \mathbf{u}^\prime_i,w_i)$ determines the weight for the $\mathbf{u}_j$ $j$th grid point. For each frequency channel there are $8128$ baselines. Each baseline has a complex conjugate pair where $\mathcal{V}(\mathbf{u}) = \mathcal{V}^{\dagger}(-\mathbf{u})$, for a total of $16256$ simulated visibilities for each frequency. To minimise computation we use a coarse kernel size of $(91\times91)$ pixels, where each pixel has size $0.5\,\lambda$. The number of operations per baseline is proportional to $N^2$, however the accuracy of the sampling kernel is asymptotic. This is a reasonable trade-off in accuracy for computational efficiency. 

Once the visibilities have been sampled, Gaussian thermal noise is added using the radiometer equation for a single baseline (see the appendix section \ref{sec:thermal-noise}). The noise level for a given baseline is determined by the observing frequency, the channel width $(\Delta\nu_f)$ and the observation time length $\Delta t$. In this work $\Delta t$ was used to control the noise level; we set $\Delta t =10^4\,\rm{hours}$ to ensure that the 21\,$\rm{cm}$ signal could be detected in a single snapshot observation. This allows for a quantitative analysis of our ability to detect the 21\,$\rm{cm}$ EoR signal in the presence of the intervening extended foreground objects. A more realistic approach would be to simulate the full $10^4\,\rm{hours}$ of observations incorporating rotation synthesis. This is however computationally expensive, and this level of complexity is not required to answer the underlying question in this paper. We will further discuss rotation synthesis in Sections \ref{sec:sidelobe-null} and \ref{sec:future-pipeline}.

\subsection{Gridding}\label{sec:gridding}

Gridding is the process by which the Fourier sky-cube is reconstructed from the visibilities; this is the first step in calculating the power spectrum. Gridding reconstructs the Fourier sky-cube as a function of frequency, by distributing the frequency dependent measured visibilities onto the $(u,v)$ plane via a smooth gridding kernel. This is important because the contributions to a single visibility come from a region of the $(u,v)$ space. Each grid point $\mathbf{u}_j$ is the weighted average of all the baselines $\mathcal{V}(\mathbf{u}_i)$ multiplied by some weight $W(\mathbf{u}_j - \mathbf{u}_i)$ determined at the $j$th grid point via
\begin{equation}\label{eq:vis-grid}
    \tilde{\mathcal{I}}(\mathbf{u}_j,\nu) = \frac{\mathlarger{\sum}\limits_{i=0}^{N_{bl}}W(\mathbf{u}_j - \mathbf{u}_i) \mathcal{V}(\mathbf{u}_i,\nu)}{\mathlarger{\sum}\limits_{i=0}^{N_{bl}} W(\mathbf{u}_j - \mathbf{u}_i)} .
\end{equation}

The weights are determined by a smooth tapered gridding kernel function. In this work we use a Gaussian kernel defined as
\begin{equation}\label{eq:Gaussian-kernel}
    W(\mathbf{u}_j - \mathbf{u}_i) = \frac{1}{2\pi\sigma^2} \exp{ \Big\{ -\frac{|\mathbf{u}_j - \mathbf{u}_i |^2}{2\sigma^2} \Big\}}.
\end{equation}

The Gaussian kernel has a width of $\sigma=2\,\lambda$, and a kernel window pixel size of $(91\times91)$, where each pixel has size $0.5\,\lambda$. Smooth tapered gridding kernels help to reduce leakage into the Fourier $k$-modes $(|k| > 0.1\,h\,\rm{Mpc^{-1}})$ of interest for detecting the 21\,$\rm{cm}$ EoR signal. Once the Fourier sky-cube has been reconstructed via the gridding process, we perform an FFT with respect to frequency to produce the reconstructed Fourier sky-cube as a function $\eta$:

\begin{equation}\label{eq:FT-nu-to-eta}
    \Tilde{\mathcal{I}}(\mathbf{u},\eta) = \int^\infty_{-\infty} \Tilde{\mathcal{I}}(\mathbf{u},\nu) e^{-2\pi i(\nu \cdot\eta) } \, d\mathbf{\nu} \:\: \rm{Jy\,Hz}.
\end{equation}

Prior to the FFT we spectrally taper the reconstructed Fourier sky-cube with a Blackman-Harris window. This tapering reduces spectral leakage introduced by aliasing from the bandwidth limited FFT in the frequency axis. Aliasing introduces a sinc function which spreads power from foreground wedge modes into higher $k_{||}$ parallel modes in the EoR window.

\subsection{Calculating the 1D and 2D Power Spectra}\label{sec:1D-2D-pspec.}
%
%
The power spectrum provides information on how Gaussian the perturbations in the 21\,$\rm{cm}$ brightness temperature are as a function of the spatial $k$-modes, which have units of $(h\rm{\,Mpc^{-1}})$ \citep{Morales2004,Furlanetto_2006}, and is the main output product of MWA EoR science \citep{MWA-sci}. The $k$-modes can be converted from the Fourier modes $(u,v,\eta)$ into the components $(k_x,k_y,k_{||})$. These conversions are outlined in \citet{Morales2004}, and are performed using Equations \ref{eq:u2k} outlined in the appendix. The power spectrum as a function of the $k$-modes is determined by averaging the product of $\Tilde{\mathcal{I}}(\mathbf{k})$ and its conjugate $\Tilde{\mathcal{I}}^\dagger(\mathbf{k})$ (denoted by the $\dagger$):

\begin{equation}\label{eq:power-analytic}
    P(\mathbf{k}) = \delta_D(\mathbf{k} - \mathbf{k}^\prime)\frac{1}{\Omega_V} \langle \Tilde{\mathcal{I}}^\dagger(\mathbf{k})\Tilde{\mathcal{I}}(\mathbf{k}) \rangle,
\end{equation} 
where $\Omega_V$ is the solid angle of the field of view; the Dirac delta $(\delta_D)$ and angular brackets represent the ensemble average over the field. Equation \ref{eq:power-analytic} is equivalent to the three dimensional Fourier transform of the two point correlation function. Due to the effective isotropy of the 21\,$\rm{cm}$ signal \citep{Furlanetto_2006}, the power spectrum represents the variance of a random Gaussian field as a function of $k$-mode. For the 1D spherically averaged power spectrum we average spherical shells:

\begin{equation}\label{eq:power-1D}
    P(\mathbf{k}) = \frac{\mathlarger{\sum}\limits_{i\in |\mathbf{k}|} \tilde{\mathcal{I}}^\dagger(\mathbf{k}_i) \tilde{\mathcal{I}}(\mathbf{k}_i) \tilde{W}(\mathbf{k}_i) }{\mathlarger{\sum}\limits_{i\in |\mathbf{k}|} \tilde{W}(\mathbf{k}_i)} \:\:\rm{Jy^2\,Hz^2}
\end{equation}
where $|\mathbf{k}| = \sqrt{k_x^2 + k_y^2 + k_{||}^2}$. The 2D cylindrically averaged power spectrum instead averages rings of $k_\perp = \sqrt{k_x^2 + k_y^2}$ as a function of $k_{||}$:

\begin{equation}\label{eq:power-2D}
    P(k_\perp,k_{||}) = \frac{\mathlarger{\sum}\limits_{i\in k_\perp} \tilde{\mathcal{I}}^\dagger(\mathbf{k}_i) \tilde{\mathcal{I}}(\mathbf{k}_i) \tilde{W}(\mathbf{k}_i) }{\mathlarger{\sum}\limits_{i\in k_\perp} \tilde{W}(\mathbf{k}_i)} \:\:\rm{Jy^2\,Hz^2}.
\end{equation}

Throughout the gridding process, the accumulated Gaussian weights for each $\mathbf{u}_j$ grid point were stored in a weights array $W(\mathbf{u})$. The new Fourier weights $\tilde{W}(\mathbf{k})$ are the frequency average of the accumulated Gaussian weights $W(\mathbf{u})$.

\subsection{The Fiducial 21cm Signal}\label{sec:21cm-signal}

For comparison with the SNR and CenA sky-model power spectra, we create noise only reconstructed Fourier sky-cube $\Tilde{\mathcal{I}}_\mathcal{N}(\mathbf{k})$ with an added fiducial simulated 21\,$\rm{cm}$ signal. Using the radiometer equation (Equation \ref{eq:radiometer} in the appendix), we generate random noise for the real and imaginary components for each visibility as a function of frequency. These visibilities are then gridded and Fourier transformed to create the noise only reconstructed Fourier sky-cube. Since the power spectrum is a measure of the variance of the underlying visibility distributions at different $k$-modes (Section \ref{sec:1D-2D-pspec.}), we use simulated models of the 21\,$\rm{cm}$ power spectrum to generate random Gaussian fields as a function of $|\mathbf{k}|$. These random Gaussian fields can then be added to $\Tilde{\mathcal{I}}_\mathcal{N}(\mathbf{k})$, approximating a full 21\,$\rm{cm}$ simulation without foregrounds. However, to properly simulate the signal we might detect with the MWA, a more accurate method would be to use a simulated 21\,$\rm{cm}$ image cube as input into the pipeline. This would capture any potential signal loss as a result of the pipeline.

In this work we use a fiducial 21\,$\rm{cm}$ power spectrum model created by \citet{21cm-FAST} using the software simulation package \textsc{21cmfast}. \textsc{21cmfast} is a semi-numerical modelling package which uses astrophysical approximations to efficiently simulate the cosmological 21\,$\rm{cm}$ signal. The generated 21\,$\rm{cm}$ power spectrum from \textsc{21cmfast} has been shown to be accurate to within $\sim10\%$ of more complex hydrodynamical numerical simulations \citep{21cm-num-sim} on spatial scales of $\geq 1\,\rm{Mpc}$.  

The fiducial 21\,$\rm{cm}$ 1D power spectrum we use in this work is calculated at a redshift of $z=7.171$ which is approximately the redshift at the centre of the simulation observing band for the EoR\,2 field $(\nu = 183\,\rm{MHz})$. The fiducial 21\,$\rm{cm}$ power spectrum is then interpolated as a function of $|\mathbf{k}|$. The interpolated power spectrum is then converted from units of $\rm{mK^2}$ to units of $\rm{Jy^2\,Hz^2}$:

\begin{equation}\label{eq:21cm-var}
    \sigma^2(\mathbf{k}) = \frac{2\pi^2}{k^3} \frac{\Delta^2(\mathbf{k})}{C} \: \:\rm{Jy^2\,Hz^2}
\end{equation}

$\Delta(\mathbf{k})$ is the power spectrum which has not been volume normalised. $C$ is a cosmological unit conversion factor which converts the power spectrum from cosmological units to $\rm{Jy^2\,Hz^2}$ (given by Equation \ref{eq:conversion-factor} in the appendix). Using Equation \ref{eq:21cm-var} and the interpolated 21\,$\rm{cm}$ power spectrum we calculate a $\sigma(\mathbf{k})$ cube for each $k$-mode, using the $k$-mode grid corresponding to the simulated visibilities. These sigma values are then used to sample a random normal distribution for both the real and imaginary components of the complex visibility. The resulting random Gaussian complex cube is our 21\,$\rm{cm}$ Fourier sky-cube as a function of $k$-modes which can be added to $\Tilde{\mathcal{I}}_\mathcal{N}(\mathbf{k})$.

\begin{figure}
    \centering
    \includegraphics[width=0.475\textwidth]{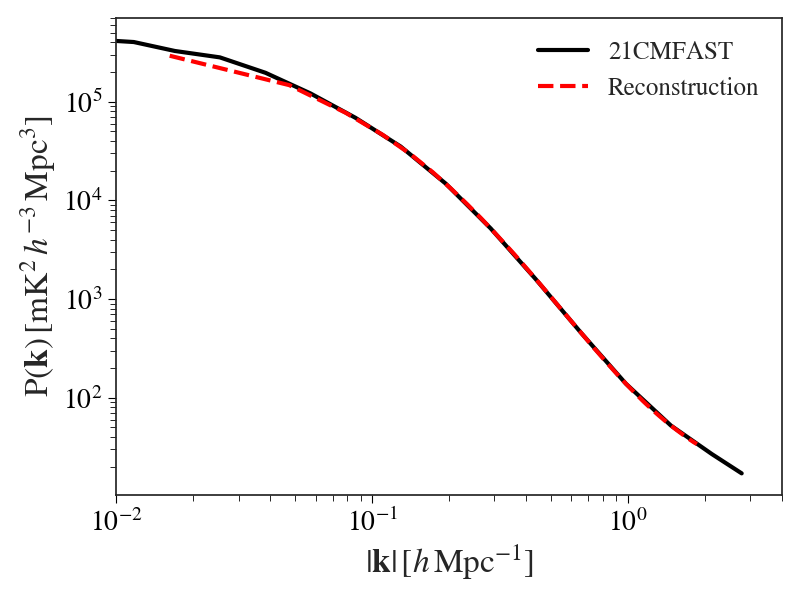}
    \caption{Comparison of the sampled fiducial signal (solid black line), and the reconstructed wedge cut fiducial signal from a spherically averaged 1D power spectrum (dashed red line). Both lines are in good agreement except at low $k$ modes. There are less samples in these modes.}
    \label{fig:1D-21cm-pspec}
\end{figure}

To test whether the noise plus random Gaussian 21\,$\rm{cm}$ Fourier sky-cube with the gridded Gaussian weights generates the expected power spectrum, we calculate the spherically averaged 1D power spectrum. Figure \ref{fig:1D-21cm-pspec} shows the fiducial 1D power spectrum signal in black, and the expected 21\,$\rm{cm}$ signal in the dashed red line. Only at the lowest $k$-modes do we not fully retrieve the expected signal, due to the relatively poor sampling at the shortest ($<100$ baselines below $k \sim 0.01\,h\rm{\,Mpc^{-1}}$) baselines.

\section{Data \& Morphological Models}\label{sec:data+models}
%
%
%
Extended radio sources such as CenA typically have angular sizes on the order of $\sim1^\circ$ or larger. Most extended radio source modelling tools such as \textsc{PyBDSF} \citep{PyBDSF}, primarily use generalised 2D Gaussian functions to fit source flux density at different angular scales. 2D Gaussian functions have great utility because they have analytical Fourier transforms, and require less components than Dirac delta models, which essentially model each pixel as an independent radio source. In this work we similarly use generalised 2D Gaussians defined below:

\begin{equation}\label{eq:2D-gauss}
    G(x,y) = G_0 e^{-\left(a(x - x_0)^2 + 2b(x - x_0)(y - y_0) +c(y - y_0)^2 \right)  }
\end{equation}
where $a$, $b$, and $c$ are parameters that simplify the expression:

\begin{align}\label{eq:2D-gauss-co}
    a &= \frac{\cos^2{\theta_p}}{2\sigma^2_x} + \frac{\sin^2{\theta_p}}{2\sigma^2_y}\\
    b &= -\frac{\sin{2\theta_p}}{4\sigma^2_x} + \frac{\sin{2\theta_p}}{4\sigma^2_y}\\
    c &= \frac{\sin^2{\theta_p}}{2\sigma^2_x} + \frac{\cos^2{\theta_p}}{2\sigma^2_y}
\end{align}
$x_0$ and $y_0$ are the $x$-axis and $y$-axis positions of the centre of the Gaussian, $\theta_p$ is the position angle the Major axis of the Gaussian makes relative to $y$-axis. $\sigma_x$ is the Gaussian width in the $x$-axis, and $\sigma_y$ is the Gaussian width in the $y$-axis. 

To correctly model the different angular scales of morphological features, we can construct a function which is a summation of Gaussians of varying sizes for the different angular scales:

\begin{equation}\label{eq:Multi-comp-Gaussian}
    I_\textrm{Source}(x,y;\hat{\theta}) = \sum^{N_\textrm{gauss}}_{i=0} G(x,y;\hat{\theta}_i)
\end{equation}

In this instance $\hat{\theta}_i = (x_0,y_0,\sigma_x,\sigma_y,\theta_p,G_0)_i$ is the vector of parameters for the $i$th component Gaussian, and $\hat{\theta}$ represents the matrix of vectors with $(N_\textrm{gauss}\times6)$ elements. To fit the multi-component Gaussian model we minimise the square residuals $(I_\textrm{Source}(x,y;\hat{\theta}) - I_\textrm{data})^2$, with the Python package \texttt{scipy} \citep{scipy}. This method performs well if the boundary conditions for the parameter space and the initial conditions are chosen well. Peak detection methods (discussed further in Sections \ref{sec:CenA-morph_model} and \ref{sec:SNR-models}), instrumental resolution, and known source sizes help to restrict the total number of components, as well as provide good initial guesses on the fit parameters. 

\subsection{Centaurus A}\label{sec:CenA-data}

In this work we utilise the best available MWA image of Cen A \citep{McKinley2021}, taken at $185\,\rm{MHz}$ with an observing bandwidth of $30.72\,\rm{MHz}$. \citet{McKinley2021} observed CenA using Phase I MWA data and Phase II extended MWA baseline configuration data. The final image has an rms background noise level of approximately $4\,\rm{mJy/beam}$ with a peak brightness of $202\,\rm{Jy/beam}$ in the inner lobes, giving the image a dynamic range of approximately $50000$. This image is free of significant artefacts, and provides the most accurate detailed representation of CenA at these radio frequencies to date \citep{McKinley2021}.

\subsubsection{Centaurus A Morphological Model}\label{sec:CenA-morph_model}
\begin{figure*}
            \centering 
            \includegraphics[width=\textwidth]{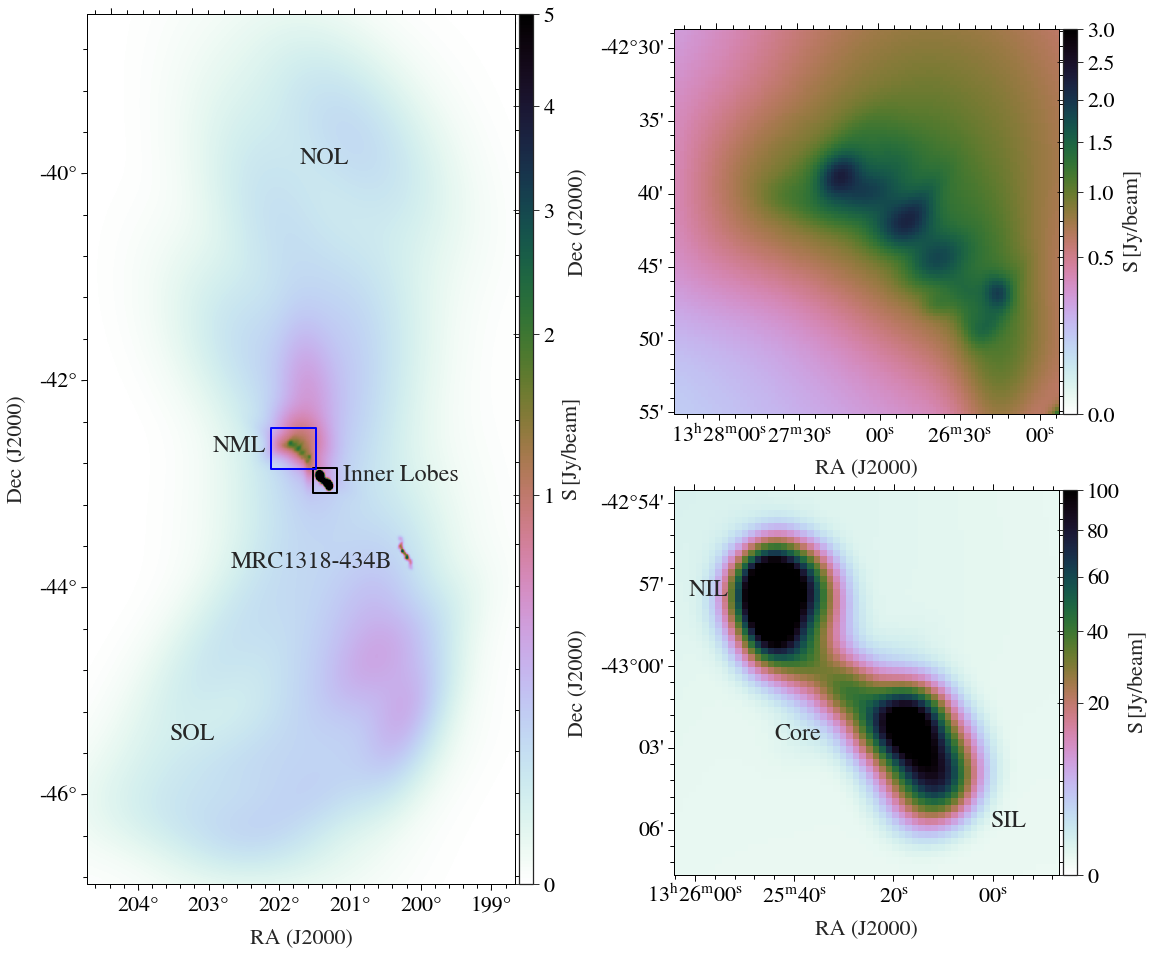}
            \caption{Model SIN projected image of Centaurus A fit to the deep multi-scale image from \citep{McKinley2021}. The leftmost panel shows the full extent of Cen A, with the different morphological regions labelled. The inner lobes and the NML in particular are shown in the solid coloured boxes. The top rightmost panel is a close up image of the inner lobes, where the Northern inner lobe (NIL) and the Southern inner lobe (SIL) are labelled separately. The bottom rightmost panel is a close up image of the NML model. The background galaxy MRC1318-434B is shown in the SOL.}
            \label{fig:CenA-model}
\end{figure*}

The image in Figure 1 from \citet{McKinley2021} was used to create a bespoke morphological model of CenA, by fitting 2D Gaussians to prominent peaks. Since this image is large ($1258\times2452$ pixels), and has four orders of magnitude in dynamic range, it was split into different regions which could be individually modelled. In particular the compact smaller scale structures of CenA such as the inner lobes, the Northern Middle Lobe (NML), and the background galaxy MRC1318-434B were separated into different images. 

%
%
The bespoke fitting process for these three images was the same; we used the Python package \texttt{skimage} to perform local peak detection with the function \texttt{peak\_local\_max} \citep{skimage}. The peak detection parameters were manually adjusted to choose an appropriate number of peaks for each image. An estimate of the appropriate number of peaks was determined by looking at the images with overlaid contours. We then used the flood fill algorithm from \texttt{skimage} to create a cutout island. Islands are subsets of the data on which 2D Gaussian fitting is performed. The flood fill parameters were manually fine tuned until the diffuse emission of each feature was almost entirely encapsulated. Peaks that lay outside of the island were removed. For the inner lobes image we identified $17$ peaks, $15$ for the NMLs image, and $12$ for the background galaxy image. 

%
%
Before each image was fitted, the background flux density was estimated by calculating the median pixel value of all the pixels outside of the island mask. The background was assumed to be constant throughout each image. This median background flux was then subtracted from the island removing the flux density offset introduced by the outer lobes of CenA. Using the island mask and peak locations, we then fitted the $N$ number of 2D Gaussians to each image using the \texttt{scipy.optimize} function \texttt{curve\_fit()} \citep{scipy}. We restricted the minimum Gaussian size to have the same parameters as the Gaussian restoring beam for the image \citep{McKinley2021}. The resulting fit for the inner lobes can be seen on the top right hand panel of Figure \ref{fig:CenA-model}, and the resulting fit to the NML can be seen on the bottom right hand panel of Figure \ref{fig:CenA-model}.

%
%
Once successful fits to the image were obtained, the models were subtracted from the main CenA image. The source finding algorithm \texttt{Aegean} (see \citet{Aegean} and \citet{Aegean2} for details) was then applied to the residual image to identify point sources that might be present in the outer lobes and the periphery. $1034$ points sources were found and subtracted from the residual CenA image; most of these sources fell outside of the outer lobes due to the lower background flux density. With the new residual image we used the \texttt{astropy} function \texttt{block\_reduce} to down sample the image by a scale factor of $19$. The reduction of the residual image scale reduces overall computational load. The new image had angular pixel sizes of $\sim5\,\rm{arcmins}$. The function \texttt{block\_reduce} can conserve the summation of the flux density in the down sampling process, which we use here. The Northern Outer Lobe (NOL) and the Southern Outer Lobe (SOL) were then separated into two cropped images, and the same source finding and fitting process applied to the inner lobes and NML was applied to the reduced outer lobe images. In total $9$ peaks were found for the NOL, approximately half of which corresponded to the large scale diffuse emission from the NML. A total of $8$ peaks were identified for the SOL. The Gaussian fits to these peaks were not restricted to a minimum size, since the pixel size is larger than the PSF in the down sampled images. A total of $61$ Gaussians (including the $12$ from the background galaxy) were fitted to the CenA image, ranging in size from the the Gaussian restoring beam PSF to $\sim2^\circ$. 

The total CenA model image can be seen in Figure \ref{fig:CenA-model}, which is separated into three panels. The large left hand side panel illustrates the entire 61 component CenA model, with the main features such as the inner lobes and the outer lobes labelled. The smaller right hand side panels illustrate the compact models of the inner lobes\footnote{The inner lobes are often separated into the Northern Inner Lobes (NIL) and the Southern Inner Lobes (SIL). This convention is retained in the top rightmost pannel.} and the NML respectively. The main morphological features are labelled in black text.

\subsubsection{Centaurus A Spectral Model}\label{sec:CenA-SI_model}
%
%
%
%
In addition to the morphology of CenA we require a spectral model at low radio frequencies to capture the spectral structure of CenA in the power spectrum. For this purpose we use the spectral index map shown in Figure 4 of \citet{Ben2018} as a guide. The spectral index distribution of CenA has been thoroughly investigated in the literature \citet{Alvarez2000,Ben2013,Ben2018}. At low radio frequencies the spectral index distribution of CenA is relatively uniform with a spectral index range of $-0.5$ to $-0.8$, and an average spectral index of $\sim-0.7$ across the entire source. There is small scale regional variation, particularly at the edge of the outer lobes and in the inner lobes \citep{Ben2018}. For this work following the suggestions from McKinley (private communication) we assign a flatter spectral index of $\alpha=-0.5$ to the inner lobes, and we assign the rest of CenA an approximate median spectral index of $\alpha=-0.7$. For the purposes of this work, a relatively simple spectral behaviour is adequate.

%
%
Using the spectral index and the derived flux density for each component of the CenA model, we compare the total integrated flux density from our CenA model\footnote{This is including the background galaxy as the comparison is made to measurements made at low resolution which confuse the background galaxy with the diffuse emission of the outer lobes.} to the measured total integrated flux density from the literature \citep{Alvarez2000,Ben2013}. We rescaled literature flux densities by a spectral index of $\alpha=-0.7$ to a frequency of $\nu=184.95\,\rm{MHz}$. The total integrated model CenA flux density is $4096 \pm 274\,\textrm{Jy}$ compared to $5538.8 \pm 817.8\,\textrm{Jy}$ for \citet{Alvarez2000}, and $4832\pm1066\,\textrm{Jy}$ for \citet{Ben2013}. The model recovers most of the flux density, with some flux density missing on intermediate and small scales in the outer lobes. The difference of $\Delta S_\textrm{tot} \pm \sim15\%$ with our model compared to \citet{Ben2013} does not affect our ability to answer the question as to whether or not CenA causes leakage into the EoR window for EoR\,2 observations. Additionally the flux scale uncertainty for the total CenA flux density calculated by \citet{Ben2013} were $\sim20\%$, so for all applied purposes in this paper the model CenA flux scale is adequate.

\subsection{Supernova Remnants}\label{sec:SNR}
%
%
The cataclysmic end to a massive star's life ejects material at high speeds into the surrounding inter stellar medium. Relativistic electrons accelerated at the shock boundaries of SNRs produce synchrotron radiation as they interact with the local magnetic field \citep{SNR-synchrotron}. This emission is dominant at radio wavelengths particularly around $1\,$GHz \citep{Stafford_2019}. Known Galactic SNRs in the low frequency radio regime have been extensively studied \citep[see][for a review]{SNR-review-2015}, and have been collated into a comprehensive catalogue \citep{Green-SNRcat}. This catalogue provides information about the position in RA and DEC, as well the major and minor elliptical sizes of each SNR. Additionally the catalogue provides the expected 1\,GHz flux density and spectral index derived from the literature where possible \citep[see][for references]{Green-SNRcat}. 

\citet{Green-SNRcat} SNR catalogue contains $294$ Galactic SNRs, $269$ of which have 1\,GHz flux density values. $25$ SNRs either had no 1\,GHz flux density estimates, or only had upper limits, and where removed from the catalogue. Of the remaining $269$ sources only $218$ had spectral index measurements, some of which are dubious \citep{Green-SNRcat}. For the $51$ SNRs that did not have spectral index values they were assigned the population median spectral index value of $\alpha \sim-0.5$ as a placeholder. The SNR flux densities were then scaled from 1\,GHz flux to a frequency of $183\,\textrm{MHz}$, which corresponds to the frequency at the centre of the simulated EoR\,2 field observations.

Further subsetting of the SNR catalogue is performed using major axis size of the remaining SNRs. A cutoff size of $\geq23\,\textrm{arcminutes}$ is applied since this is twice the size of the $>300\lambda$ ($\sim11.5\,\textrm{arcminutes}$) $uv$-cutoff. This cutoff is applied in $uv$-space to the visibilities because the 21\,$\rm{cm}$ signal power is expected to be the greatest at larger spatial scales \citep{Furlanetto_2006}. After applying the major-axis size condition, the SNR catalogue only has $101$ remaining SNRs. Additional subsetting is performed for SNRs below a declination of $+30\deg$, of which there are $73$. Sources above this cutoff are not contained in The GaLactic and Extra-galactic All-sky Murchison Widefield Array (GLEAM) survey. GLEAM was an all sky survey that observed the southern sky below declinations of $+30\deg$ using the MWA \citep{GLEAM}, images from GLEAM are publicly available through the GLEAM VO server \citep{GLEAMyr1}\footnote{\url{http://gleam-vo.icrar.org/gleam_postage/q/form}}. For each of these sources we download $200\,$MHz cutout images from the GLEAM VO server. The $200$\,MHz wideband GLEAM image is the most sensitive with an angular resolution of $\sim2\,\textrm{arcminutes}$ \citep{GLEAMyr1}. The 2D Gaussian component fitting to these images is described in the following section.

\begin{figure*}
    \begin{center}
        \begin{subfigure}[b]{0.495\textwidth}
            \centering
            \includegraphics[width=\textwidth]{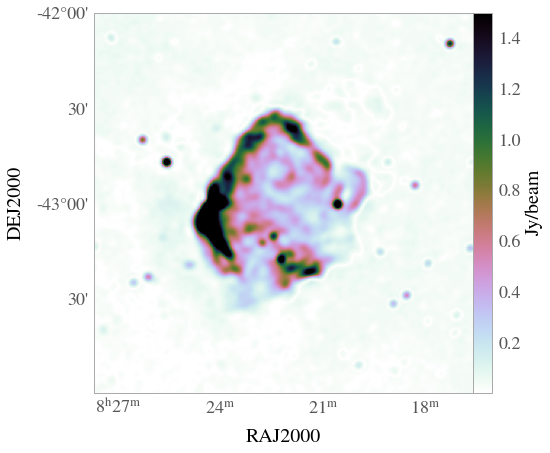}
            \caption[]%
            {{\small Puppis A 200\,MHz GLEAM Image}}
            \label{fig:2a}
        \end{subfigure}
        \hfill
        \begin{subfigure}[b]{0.495\textwidth}
            \centering
            \includegraphics[width=\textwidth]{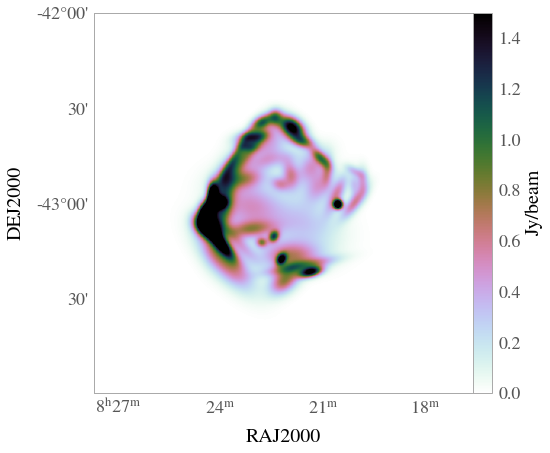}
            \caption[]%
            {{\small 41 Component Puppis A Gaussian Model Image}}
            \label{fig:2b}
        \end{subfigure}
        \begin{subfigure}[b]{0.495\textwidth}
            \centering
            \includegraphics[width=\textwidth]{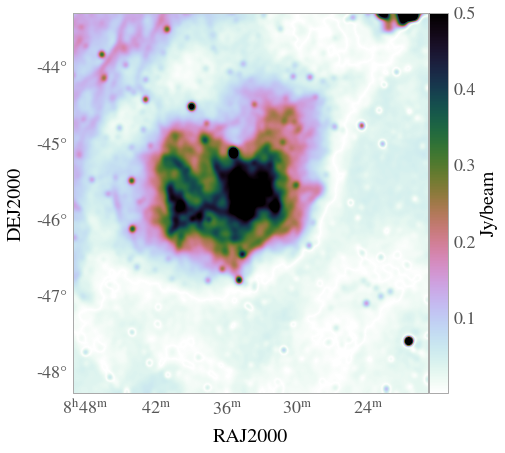}
            \caption[]%
            {{\small Vela 200\,MHz GLEAM Image}}
            \label{fig:2c}
        \end{subfigure}
        \begin{subfigure}[b]{0.495\textwidth}
            \centering
            \includegraphics[width=\textwidth]{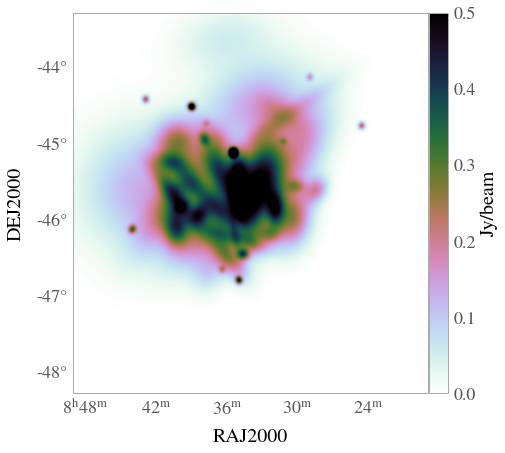}
            \caption[]%
            {{\small 42 Component Vela Gaussian Model Image}}
            \label{fig:2d}
        \end{subfigure}
        \caption[]
        {\small GLEAM cutout images of Puppis A (Subfigure (a)) and Vela (Subfigure (b)) at 200\,MHz.
        The Puppis A image has a peak of $4.50\,[\textrm{Jy/beam}]$, and is convolved with a Gaussian restoring beam with a major and minor size of $a=2.23\,[\textrm{arcmin}]$, $b=2.14\,[\textrm{arcmin}]$, and a position angle of $\sim315^\circ$ relative to North. The Vela image has a peak flux density of $1.62\,[\textrm{Jy/beam}]$, and is convolved with a Gaussian restoring beam with a major and minor size of $a=2.23\,[\textrm{arcmin}]$, $b=2.14\,[\textrm{arcmin}]$, and a position angle of $\sim351^\circ$ relative to North. Due to the size $(5\times5 \deg)$ of the Vela image, it is further convolved with a Gaussian of size $a_\textrm{maj}\sim5.41\,\textrm{[arcmin]}$. The lower resolution allowed for the fit of fewer components to the Vela image. This does not affect the sky-models in this work, since the sky-model image cube resolution is $\sim11\,\textrm{arcmin}$. Subfigure (b) is the 41 component Gaussian model for Puppis A, fit only to an $1^\circ$ circular cutout. The Puppis A model image has a peak flux density of $4.48\,[\textrm{Jy/beam}]$. Subfigure (c) is the 42 component model of Vela, which was fit to a $\sim4^\circ$ circular cutout of Vela, and has a peak flux density of $1.60\,[\textrm{Jy/beam}]$.} 
        \label{fig:SNR-models}
    \end{center}
\end{figure*}

\subsubsection{SNR Morphological Models}\label{sec:SNR-models}
%
%
For some SNRs which have relatively low surface brightness, island fitting methods such as \texttt{Aegean} and PyBDSF \citep{PyBDSF} have a tendency to over-fit the wide-band $200\,\rm{MHz}$ GLEAM cutout image backgrounds. Due to the relatively large number of GLEAM cutout images $(N=73)$, we instead opted to develop an automated fitting method which utilises prior information about the size, and location of each SNR. The prior information is taken from the SNR catalogue, where the major axis and the centroid RA and DEC position for the SNR is used to create an island mask. 

%
%
The fitting method employed to fit each SNR was similar to the bespoke method developed for CenA, with some key differences. In particular we took a more accurate approach in calculating the image background. This is particularly important for SNRs that have a low surface brightness compared to the image background. The GLEAM SNR cutout images do not have the large dynamic range of the CenA image from \citep{McKinley2021}. For the SNRs the background emission was determined through an iterative approach, where the pixels outside the island where averaged. The fitting algorithm then calculates the root mean squared (rms) of the masked image (island pixels set to NaN). We use a default rms threshold of $2.5\sigma$ above the median background to mask potential point sources. The median background and rms are then recalculated and further thresholding performed. Convergence to a single background noise level for each cutout image was quick, typically taking a max number of five iterations, this was set as the default.

%
%
Once the background has been calculated it is subtracted from the island image. We then perform peak detection using the \texttt{skimage} function \texttt{blob\_dog()}. This method blurs the image with increasing standard deviations (in terms of pixel coordinates), and calculates the difference between successive images which are then stacked into a difference image cube. Blobs or peaks are identified as local maximums in the data cube. This allows for the detection of different scales of peaks \citep{skimage}. 

%
%
After peak detection, we then fit 2D Gaussians using the \texttt{scipy.optimize} function \texttt{curve\_fit()}, as we did when fitting CenA. The fitting parameter space is restricted by constraining the maximum Gaussian fit size to a fraction of the SNR major axis (default fraction is $1/8$)\footnote{The $1/8$ size constraint was found to be reasonable, since most observed SNR morphologies are generally dominated by smaller scale filament like structures \citep{SNR-review-2015}}. The fitting space is also restricted to be within the island, minimising spurious fits outside the island. Additionally, the minimum 2D Gaussian size is restricted to match the image restoring beam.

%
%
To test the validity of the multi-component fit model, we also fit a single 2D Gaussian to each SNR image. For some filled type SNRs this model might be a more accurate representation of the morphology, additionally allowing for an automated comparison which can distinguish between potentially real and spurious fits. However, many fits still had to be assessed by eye to ensure the multi-component models were not fitting noise, or image artefacts. The single 2D Gaussian fit only has two free parameters, the Gaussian amplitude and the position angle. The Major and Minor axis sizes are fixed from the information from SNR catalogue. To compare the multi-component fits to the single Gaussian fit we utilise the Bayesian Information Criterion (BIC) \citep{BIC}:

\begin{equation}
    \textrm{BIC} = \chi^2 + k\log{n},
\end{equation}
where $\chi^2$ is the sum of the squared residuals which have been normalised by the squared image rms, $k$ is the number of model fit parameters, and $n$ is the number of data points. The model with the lower $\textrm{BIC}$ is the preferred fit \citep{BIC}, which for most SNRs is typically a multi-component model. Some sources were too faint to be present in the GLEAM $200\,$MHz images, and peaks were fit to sidelobe confusion noise, or to artefacts. In these cases we replaced these fits with the single Gaussian fit. In total out of the $73$ fit candidates $24$ had a preferential single Gaussian fit. 

To determine the accuracy of the SNR models to the expected flux density, the total integrated model flux density for each SNR was compared to the expected flux density provided by \citet{Green-SNRcat}. The median ratio for all SNRs was $\sim1.1\pm0.4$, with one outlier the Vela SNR model having a ratio of $17.9$. The expected flux density for Vela as quoted in \citet{Green-SNRcat} was determined from single dish Parkes observations made by \citet{Vela-II}. The GLEAM images are missing baselines below $60\,\textrm{m}$ and thus large scale flux density from Vela. 

Figure \ref{fig:SNR-models} shows example fit models of Puppis A, and Vela compared to their corresponding GLEAM images. The left hand panels are the original GLEAM images, with Puppis A on the top row and Vela on the bottom row. The model images are on the right hand side with Puppis A on top row and Vela on the bottom row. 


\subsection{Constructing Sky-Models}
%
%
%
%
%
%
%
%
%
The model fit parameters for CenA and the Galactic Plane SNRs were collated into a FITS table which contains the RA, DEC, spectral index, the total model integrated $200$\,MHz flux density, the major and minor axes, as well as the position angle for each component. Using this table, models of the entire sky in image space can be generated. For a single frequency slice the sky-model image array can be described as the aggregate of all of the model sources:

\begin{equation}\label{eq:sky-model}
    I_\textrm{sky}(\mathbf{l}) = \sum^{N_\textrm{source}}_{i=1} I_\textrm{source,i}(\mathbf{l})
\end{equation}

This aggregate modelling approach is useful, because it allows for the creation of partial sky-models, effectively simulating source subtraction. This can be used to determine how much of the Galactic Plane SNRs and CenA need to be removed in order to retrieve the 21\,$\rm{cm}$ signal in the power spectrum. For a given observation time, we calculate the Azimuth and Altitude for each source and its model components using \texttt{astropy} \citep{Astropy-I,Astropy-II}. Sources which are below the observation horizon ($\theta_\textrm{alt} < 0$) are ignored. Substituting Equation \ref{eq:Multi-comp-Gaussian} into Equation \ref{eq:sky-model} generalises the description of the total sky model to the aggregate of all the model 2D Gaussian components:

\begin{equation}\label{eq:sky-model-2}
    I_\textrm{sky}(\mathbf{l}) = \sum^{N_\textrm{source}}_{i=1} \sum^{N_\textrm{i,gauss}}_{j=1} G_j(\mathbf{l};\hat{\theta}_j),
\end{equation}
where the $j$th source has $N_\textrm{j,gauss}$ Gaussian components, with each component having $\hat{\theta}_j$ model parameters. For a zenith phase centre, the $(l,m)$ plane is an orthographic projection. Due to the small angle approximation, the Gaussian models were defined in a 2D plane, however when placing them in the $(l,m)$ frame they will need to be correctly projected. The Major and Minor axes for all Gaussians are recalculated as a function of their Altitude angle. This conserves the total flux density of the source. The projection effect is continuous, however to simplify calculations we use an approximation. For more details on how the projection is calculated refer to the Appendix Section \ref{sec:app-gauss-proj}.

The OSIRIS pipeline accepts a sky-model cube $I(\mathbf{l},\nu)$ which varies as a function of frequency. In this work we assume that the source morphology does not evolve with frequency across the simulated observation bandwidth $(15.36\,\rm{MHz})$. This is a reasonable assumption since we fit wideband images $(\geq 30.72\,\rm{MHz})$ of SNRs and CenA. We also assume that the spectral behaviour of the source components can be modelled with a power law $I \propto \nu^\alpha$, where $\alpha$ is the spectral index. This simplifies the calculation of the sky-model cube, since the OSIRIS pipeline only calculates a template Gaussian which can be scaled as a function of frequency. The iterative sum for each Gaussian model component $j$ for the $i$th source is described below:

\begin{equation}\label{eq:sky-model-3}
    I_\textrm{sky}(\mathbf{l},\nu) = \sum^{N_\textrm{source}}_{i=1} \left( \frac{\nu}{\nu_0} \right)^{\alpha_i} \sum^{N_\textrm{i,gauss}}_{j=1} G_j(\mathbf{l};\hat{\theta}_j).
\end{equation}

Some Gaussians have $\sigma << \Delta l$ (pixel size), and therefore are not properly sampled by the coarse pixel grid. One solution is to increase the grid size to effectively sample the smallest Gaussian model, however this drastically increases the required computational resources. Furthermore, we are not interested angular scales less than $\sim10\,\rm{arcmin}$. Instead we set the minimum angular Major and Minor axis size to be equal to the pixel size (which is $\sim8.4\,\rm{arcmin}$), which conserves flux density and effectively sets these smaller components as point sources.

\begin{figure*}
    \begin{center}
        \begin{subfigure}[b]{0.495\textwidth}
            \centering
            \includegraphics[width=\textwidth]{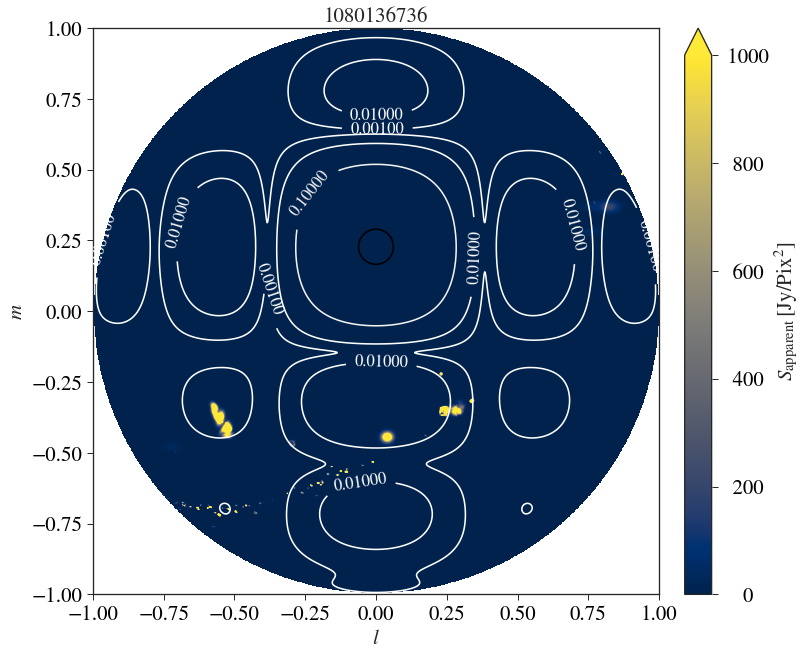}
            \caption[]%
            {{\small Sidelobe Sky-model}}
            \label{fig:sky-mod1}
        \end{subfigure}
        \hfill
        \begin{subfigure}[b]{0.495\textwidth}
            \centering
            \includegraphics[width=\textwidth]{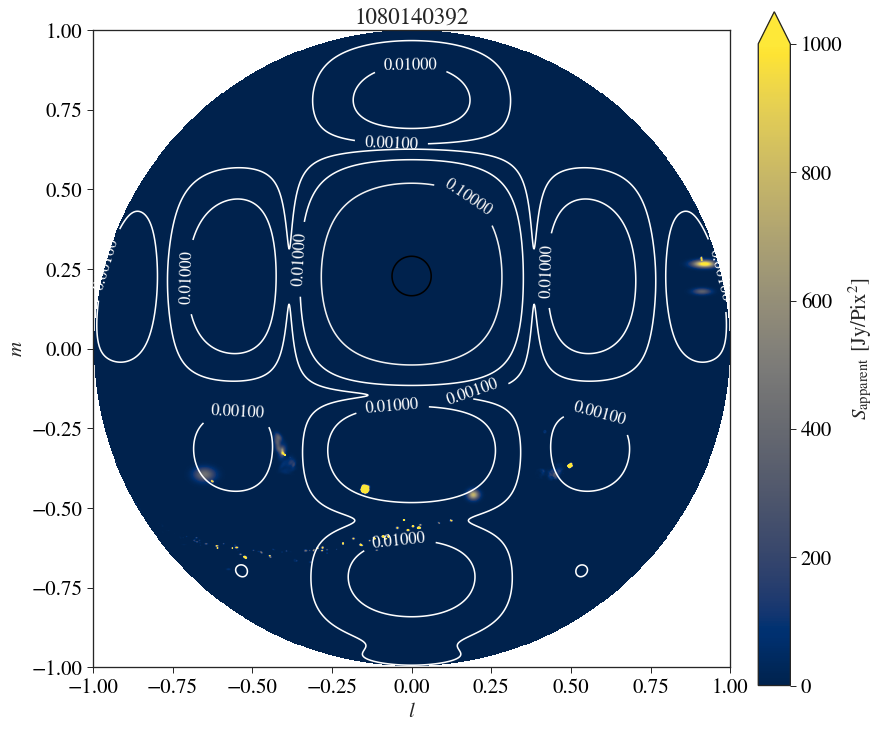}
            \caption[]%
            {{\small Null Sky-model}}
            \label{fig:sky-mod2}
        \end{subfigure}
        \caption[]
        {\small Example apparent sky-model images for sidelobe sky-model (Subfigure (a)) and the null sky-model (Subfigure (b)). The solid white contours show the MWA primary beam with lines at levels $[10^{-3},10^{-2},10^{-1},0.9]$. In Subfigure (a) Centaurus A can be seen in a sidelobe at $l\sim-0.5$ and $m\sim-0.35$. The Galactic Plane SNR sources can be seen in an arc intersecting several sidelobes, Vela and Puppis A are both visible at $l\sim0.25$, and $m\sim-0.35$. In Subfigure (b) Centaurus A and Vela have migrated out of their respective sidelobes and into primary beam nulls. Puppis A in Subfigure (b) has migrated into another sidelobe.} 
        \label{fig:sky-mods}
    \end{center}
\end{figure*}

\section{Results}\label{sec:results}
%
%
%
%
%
%

EoR\,2 field MWA observations have CenA positioned in one of the MWA primary beam sidelobes, which is a concern for EoR science. The contribution of CenA to the power spectrum is expected to be greater than the 21\,$\rm{cm}$ signal on degree size scales that are important for EoR science. Rotation synthesis will mitigate some of the power of CenA as it rotates from the sidelobe into a primary beam null. However a full simulation of hundreds of hours of MWA observations for the EoR\,2 field is not necessary to determine whether CenA and Galactic Plane SNRs introduce leakage into the EoR window. Therefore the OSIRIS pipeline only simulates a single time step, and thus does not incorporate rotation synthesis. In conjunction with CenA a procession of Galactic Plane SNRs rotates through one of the MWA primary beam sidelobes for the EoR\,2 field. The aggregate power of the Galactic Plane SNRs will not be as strongly affected by rotation synthesis, but will however vary as different sources become more prominent. Equation \ref{eq:sky-model} allows for the construction of partial sky-models which simulate the subtraction of CenA and Galactic Plane SNRs. In this section we investigate the 2D and 1D power spectrum of several input sky-models of the EoR\,2 field. In particular we look at two distinct observations to analyse the different spectral characteristics, and how the resulting leakage affects the detectability of the 21\,$\rm{cm}$ signal.

\subsection{Sidelobe and Null Test Observations}\label{sec:sidelobe-null}

To characterise the effects of rotation synthesis we simulate two sky-models of the Galactic Plane and CenA separated by one hour in time. The first observation has CenA situated in a sidelobe of the MWA primary beam (herein referred to as the sidelobe observation), and the second observation has CenA situated in a null of the MWA primary beam (herein referred to as the null observation). Figures \ref{fig:sky-mod1} and \ref{fig:sky-mod2} show the average apparent sky-models across the entire observing bandwidth, where the sky-model cube was attenuated by the FEE MWA primary beam model, and averaged as a function of frequency. The average MWA primary beam pattern across the bandwidth is shown with the solid white contours. Subfigure \ref{fig:sky-mod1} shows the sidelobe sky-model with CenA clearly visible in the sidelobe. Subfigure \ref{fig:sky-mod2} shows the null sky-model with CenA rotated into the primary beam null. 

\begin{figure*}
    \begin{center}
        \begin{subfigure}[b]{0.495\textwidth}
            \centering
            \includegraphics[width=0.85\textwidth]{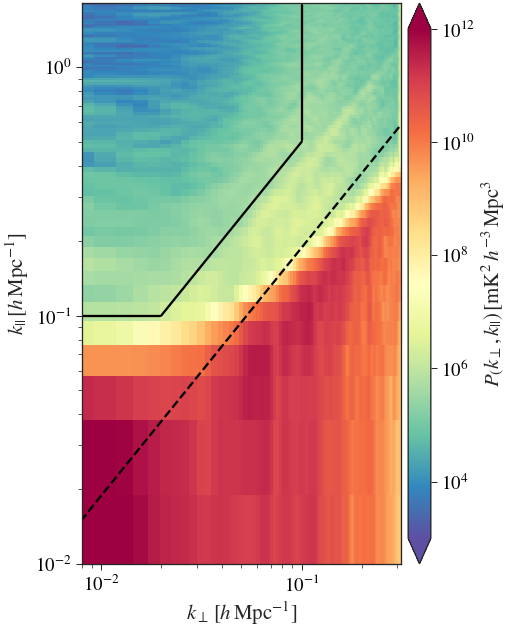}
            \caption[]%
            {{\small CenA sidelobe power spectrum}}
            \label{fig:sidelobe}
        \end{subfigure}
        \hfill
        \begin{subfigure}[b]{0.495\textwidth}
            \centering
            \includegraphics[width=0.85\textwidth]{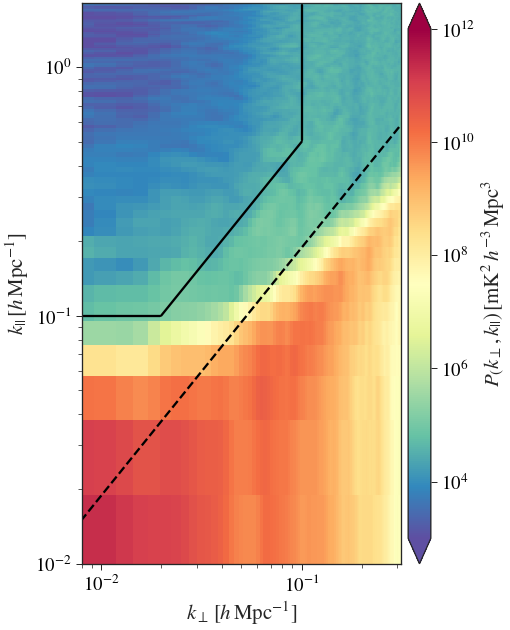}
            \caption[]%
            {{\small CenA null 2D power spectrum}}
            \label{fig:null}
        \end{subfigure}
        \begin{subfigure}[b]{0.495\textwidth}
            \centering
            \includegraphics[width=0.85\textwidth]{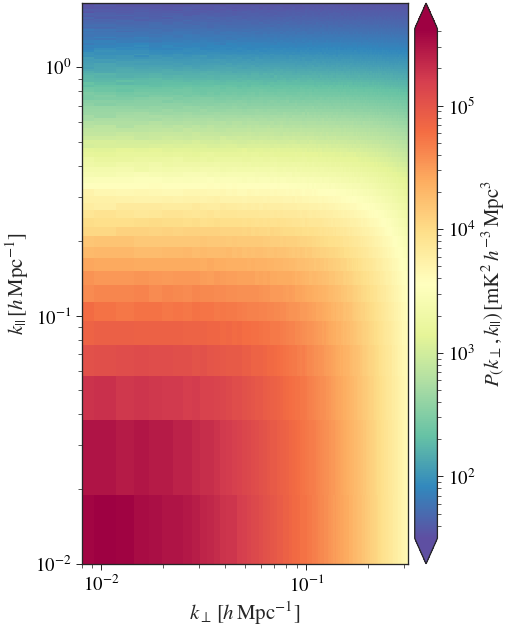}
            \caption[]%
            {{\small 21cm 2D power spectrum}}
            \label{fig:2D-21cm-power spectrum}
        \end{subfigure}
        \begin{subfigure}[b]{0.495\textwidth}
            \centering
            \includegraphics[width=0.85\textwidth]{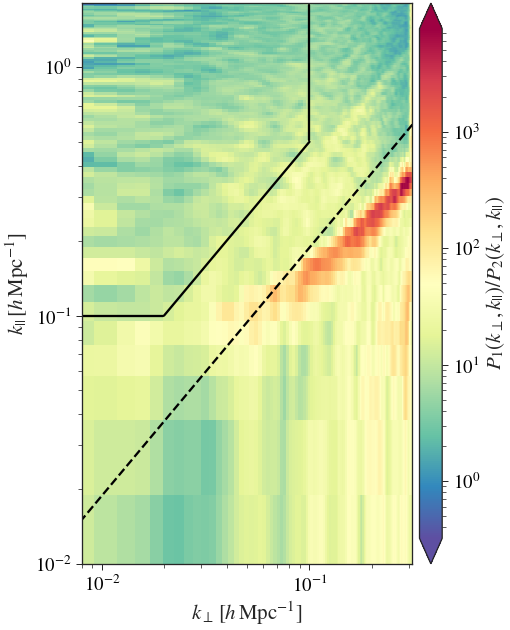}
            \caption[]%
            {{\small Sidelobe/Null 2D Power Spectrum Ratio}}
            \label{fig:Ratio}
        \end{subfigure}
        \caption[]
        {\small 2D power spectra for the sidelobe, null simulation sky-models, and the fiducial 21\,$\rm{cm}$ 2D power spectrum. Subfigure (a) is the 2D power spectrum for the sidelobe case, the solid black line indicates the wedge cut used to calculate the 1D power spectrum in Figure \ref{fig:1D-pspec}, the gradient of the solid black line indicates the horizon. The dashed black line indicates the gridding kernels field of view. Subfigure (b) is the 2D power spectrum for the null simulation. Subfigure (c) is the 2D power spectrum of the fiducial 21\,$\rm{cm}$ signal. Subfigure (d) is the ratio of the sidelobe 2D power spectrum to the null 2D power spectrum simulation. Subfigure (a) and (b) have the same colourbar scale.} 
        \label{fig:2D-pspec}
    \end{center}
\end{figure*}

In addition to the sidelobe and null observation simulations, we perform a third simulation of the sidelobe observation without CenA where the model just contains the Galactic Plane SNRs. By comparing the relative difference in the magnitude of the resulting 2D power spectrum we can determine what effect rotation synthesis may have on these observations for different $k$-modes. We can also compare this to the expected 21\,$\rm{cm}$ power expected on these modes. Figure \ref{fig:2D-pspec} shows the resulting 2D power spectrum for the sidelobe observation (\ref{fig:sidelobe}), the null observation (\ref{fig:null}), the fiducial 21\,$\rm{cm}$ 2D power spectrum (\ref{fig:2D-21cm-power spectrum}), and the ratio of the sidelobe and null 2D power spectrum (\ref{fig:Ratio}).

The solid and dashed black lines in Figure \ref{fig:2D-pspec} show the expected horizon for the entire sky, and the edge of the field of view \citep{Morales_2012,Trott_2012}. The horizon line demarcates the bright foreground wedge from the relatively clean EoR window. To assess the level of leakage we compare the average power in a small window defined by $k_\perp \in [0.01,0.03]$ and $k_{||} \in [0.1,0.3]$ for the sidelobe, null, and 21\,$\rm{cm}$ 2D power spectra. The average window power in the 21\,$\rm{cm}$ 2D power spectrum is $1.8\times10^4$\cosmounits compared to $3.44\times10^5$\cosmounits for the sidelobe 2D power spectrum, and $3.5\times10^4$\cosmounits for the null 2D power spectrum. The sidelobe observation is $\sim20$ times greater than the expected 21\,$\rm{cm}$ signal in the window, compared to a factor of $\sim2$ greater for the null observation. For comparison the average window power for a sidelobe observation which contains only CenA is $3.36\times10^5$\cosmounits, clearly showing that CenA is the dominant source of leakage for the sidelobe observation. Subfigure \ref{fig:Ratio} shows the excess power of the sidelobe observation compared to the null observation. The largest ratio values (of order 10$^3$) are mostly confined to the foreground wedge and at higher $k_\perp$, which corresponds to smaller spatial scales. The median ratio in the EoR window is $8.2$, which is indicative of the order of magnitude difference in leakage through the EoR window.

We perform a similar assessment of leakage for a single zenith flat spectrum point source, with an apparent flux density of $10.2\,\rm{Jy}$ (this is the same as CenA for the sidelobe observation). In this case we perform a noiseless simulation and remove the primary beam, only keeping the spectral tapering. The spectral tapering with the Blackman-Harris window will have sidelobes that will contribute leakage into the window. Performing the same window calculation as per the CenA simulation, we find the median power in the window for the flat spectrum source is $22.4$\cosmounits, this is $\sim3$ orders of magnitude less than the expected $21\,$cm signal. Therefore we conclude that the Blackman-Harris sidelobes are not the primary contributor to the leakage seen in the EoR window.

\begin{figure}
    \centering
    \includegraphics[width=0.45\textwidth]{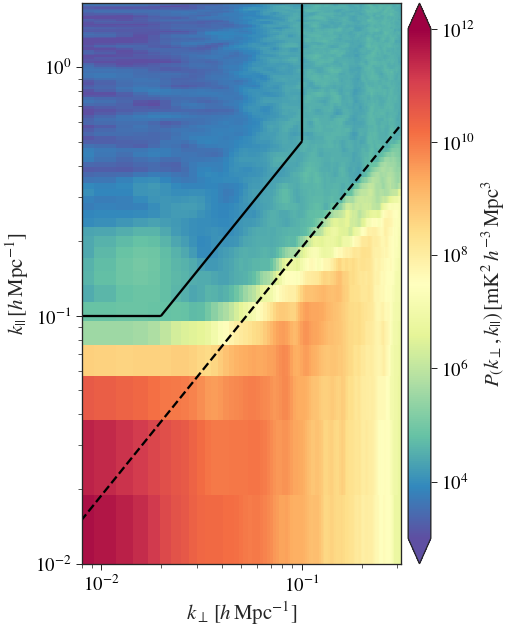}
    \caption{2D Power spectrum of the sidelobe simulation without CenA in the sky-model. The colourbar scale is the same as the 2D power spectrum in Subfigure \ref{fig:sidelobe}. There is a clear difference between this 2D power spectrum and that shown in Subfigure \ref{fig:sidelobe}, with this 2D power spectrum resembling the null 2D power spectrum.}
    \label{fig:sidelobe-CenA}
\end{figure}

We also investigated the 2D power spectrum of the sidelobe simulation without CenA in the sky-model, serving as a useful comparison to the null observation. The resulting 2D power spectrum is shown in Figure \ref{fig:sidelobe-CenA}; the colourbar is the same scale as those in Figure \ref{fig:2D-pspec}.
The average power in the window for the sidelobe minus CenA 2D power spectrum is $2.7\times10^4$\cosmounits. This is a similar level of power compared to the null observation, however the only contribution to leakage in the window is from Galactic SNRs in this case. The similarity between the null simulation and the sidelobe minus CenA simulation may indicate a potential mitigation strategy for reducing the contribution from CenA in EoR\,2 observations. However the leakage from Galactic Plane SNRs is still significant, and the change in the spectral properties and intensities of SNRs as the Galactic Plane rotates through the primary beam could be significant. 

\subsection{Partial Sky Models}\label{sec:partial-models}

\begin{figure*}
    \begin{center}
        \begin{subfigure}[b]{0.495\textwidth}
            \centering
            \includegraphics[width=\textwidth]{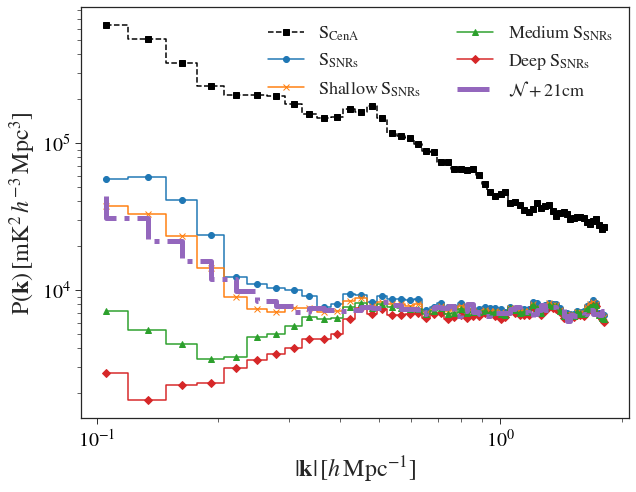}
            \caption[]%
            {{\small Sidelobe Partial 1D Power Spectra}}
            \label{fig:1D-pspec-sidelobe}
        \end{subfigure}
        \hfill
        \begin{subfigure}[b]{0.495\textwidth}
            \centering
            \includegraphics[width=\textwidth]{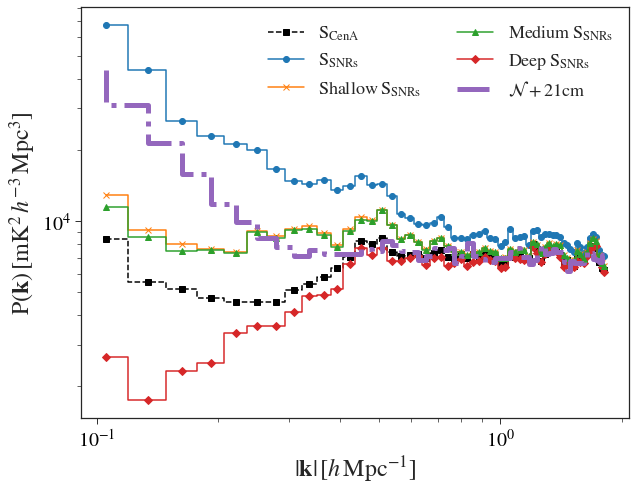}
            \caption[]%
            {{\small Null Partial 1D Power Spectra}}
            \label{fig:1D-pspec-null}
        \end{subfigure}
        \caption[]
        {\small 1D power spectra for a series of partial input sky-models, showing linearly spaced $k$ bin widths. The black square markers with the dashed black line represent the CenA only sky-model, the solid blue circles represent the total SNR sky-model. The orange crosses, the green triangles and the red diamonds are the 1D power spectra are for partial sky-models with upper limit $10\%$ (deep), $50\%$ (medium) and $90\%$ (shallow) total model apparent flux density. The actual percentages for the deep, medium and shallow partial sky-models are $\sim10\%$, $\sim36\%$ and $\sim76\%$ for the sidelobe observation, and $\sim10\%$, $\sim50\%$ and $\sim75\%$ for the null observation. The dash dot purple line with no markers is the fiducial 21\,$\rm{cm}$ signal. Subfigure (a) shows the partial sky-models for the sidelobe observation. The medium sidelobe partial sky-model is on the order of the fiducial 21\,$\rm{cm}$ power spectrum, the deep partial sky-model is below the fiducial 21\,$\rm{cm}$ power spectrum. Subfigure (b) shows the partial sky-models for the null observation. The medium partial sky-model is below the fiducial 21\,$\rm{cm}$ power spectrum, the shallow partial sky-model has a similar power to the medium partial sky-model. The similarities between the null medium and shallow partial sky-models is a result of two large bright single Gaussian sources.} 
        \label{fig:1D-pspec}
    \end{center}
\end{figure*}

%
%
%
Figure \ref{fig:sidelobe-CenA} demonstrates that even without CenA in the input sky-model, the leakage of power into the EoR window from Galactic plane SNRs is on the order of the expected fiducial 21\,$\rm{cm}$ signal power. In this section we assess how much of the SNRs need to be subtracted from the sidelobe and null sky-models in order to significantly recover the 21\,$\rm{cm}$ signal. To test this we generated a series of partial sky-model simulations for both the sidelobe and null sky-models without the 21\,$\rm{cm}$ signal. The sky-model catalogue was ordered by the apparent flux density from the faintest to the brightest source; the fractional total apparent flux density for each source was then calculated. We then generated three sky-models for each observation with upper limits of $10\%$, $50\%$ and $90\%$ of the total apparent sky-model flux density. We shall refer to these as the deep, the medium, and the shallow partial apparent sky-models respectively. The partial sky-model method assumes an ideal case where we can subtract $100\%$ of a sources total flux density. However, in reality this is not possible; simulating source subtraction errors (position or amplitude errors specifically) will not affect the main question of this paper. The partial sky-models along with the total SNR sky-model, and the CenA only sky-model for both observations were run through the OSIRIS pipeline. The 1D power spectrum was then calculated from window modes defined by $k_{||} > 0.1\,h\rm{\,Mpc^{-1}}$, $k_\perp > 0.1\,h\rm{\,Mpc^{-1}}$, and $(k_\perp,k_{||})$ modes above the horizon\footnote{The horizon $k$-mode cut is defined by the relationship: $k_{||} > \frac{\pi}{2}\frac{D_M E(z)}{D_H (1+z)} k_\perp$ \citep{Morales_2012}, where $D_M$ is the co-moving distance, $D_H$ is the Hubble distance, $\pi/2$ is the radius of the sky in radians, and the function $E(z)$ is defined by \citet{Hogg2000}.}. We also calculated the 1D power spectrum for the fiducial 21\,$\rm{cm}$ signal plus the simulation noise $(\mathcal{N})$. The resulting 1D power spectrum for both observations and the respective partial and total sky-models can be seen in Figure \ref{fig:1D-pspec}.

The orange crosses, solid green triangles, and the solid red diamonds show the deep ($90\%$), the medium ($50\%$), and the shallow ($10\%$) upper limit partial sky-model power spectrum for both the sidelobe and null observations in Figure \ref{fig:1D-pspec}. Since the partial sky-models are discretised by source and ordered from faintest to brightest, the relative percentages for the deep, medium and shallow partial sky-models are different for the sidelobe and the null observations. For the sidelobe observations the relative percentages approximately are $10\%$, $36\%$, and $76\%$ for the deep, medium and shallow partial sky-models. For the null observation the relative percentages are approximately $10\%$, $50\%$, and $74\%$ respectively. For reference, the total SNR sky-model power spectrum and the CenA only sky-model power spectrum are shown with the solid blue circles and the solid black squares respectively. The dash dot purple line is the fiducial 21\,$\rm{cm}$ signal with a $10000\,\rm{hr}$ noise level. 

%
The sidelobe and null observations have a similar total apparent brightness ($\sim8\,$Jy for both), however in Figure \ref{fig:1D-pspec} there is significant difference in the total 1D power spectrum. The null and sidelobe observations are separated by one hour in time and therefore most of the the SNRs in the model are the same, but in different parts of the MWA primary beam. For small and faint SNRs this has little impact on the power spectrum, as can be seen from the similarities in structure and power for the deep and medium upper limit partial sky-models for the sidelobe and null observations. However, this matters for the brightest most prominent sources which affect the shallow partial sky-model and the total SNR sky model. The difference between the medium, the shallow and the total 1D power spectra for both the null and the sidelobe observations are typically one or two bright extended sources; their morphology and the primary beam spectral structure imparted upon them, has the biggest impact on leakage in the 1D power spectrum.

%
For the sidelobe observation the total sky-model and the shallow partial sky-model are the same order of magnitude as the fiducial 21\,$\rm{cm}$ signal, indicating significant contamination of the signal. In contrast the null observation shallow partial sky-model is significantly below the expected 21\,$\rm{cm}$ signal on modes $|\mathbf{k}| < 0.3 \,h\rm{\,Mpc^{-1}}$. Additionally, there is little difference between the null shallow and medium partial sky-model 1D power spectra. The difference between the shallow and medium sky-models is two exceptionally large $(\sim3\,\deg)$ SNRs G$205.5+00.5$, and G$330.0+15.0$\footnote{The SNRs G$205.5+00.5$, and G$330.0+15.0$ are also known as the Monoceros Nebula, and the Lupus Loop.}. Together their apparent brightness is $\sim2\,$Jy. Due to their large degree-scale sizes these sources did not have the surface brightness to be fitted by the GLEAM cutout image method in Section \ref{sec:SNR-models}. Therefore, these sources are modelled by single component Gaussians and are missing the smaller scale structures present in their morphology. The large single Gaussian components act as a spatial filter in the $k_\perp$ axis, modulating and restricting all leakage to $k_\perp < 0.02 \,h\rm{\,Mpc^{-1}}$ modes. When calculating the 1D power spectrum, the relatively few number of modes $k_\perp < 0.02 \,h\rm{\,Mpc^{-1}}$ with significant leakage are averaged over, reducing their contribution to the window. This demonstrates the sensitivity of this type of analysis to the details of extended source morphology, and why accurate SNR subtraction is crucial. Future work will better model large single component SNRs to more accurately investigate their leakage. 

For both the sidelobe and null observations the 21\,$\rm{cm}$ signal has a power ratio of $\sim5-10$ at $|\mathbf{k}| \in [0.1,0.3] \,h\rm{\,Mpc^{-1}}$ for the deep partial sky-model. For the medium partial sky-model ($\sim50\%$) the signal to model power ratio is approximately $\sim2-3$, requiring at least $90\%$ subtraction of the SNRs from the sidelobe and null sky-models in this simulation to retrieve a significant detection of the 21\,$\rm{cm}$ signal. 

\section{Discussion}\label{sec:Discussion}

Using an input sky-model of Galactic Plane SNRs and CenA processed through an MWA simulation and power spectrum pipeline, we demonstrate that extended radio sources in the sidelobes of EoR\,2 observations introduce leakage up to an order of magnitude greater than the 21\,$\rm{cm}$ signal into the EoR 2D power spectrum window. This work shows that almost all of these widefield extended sources must be removed from the visibilities, in order to reduce contamination on EoR significant $k$-modes $(|\mathbf{k}| \lesssim 0.1\,h\,\rm{Mpc^{-1}})$, down to $\sim10-20\%$ of the 21\,$\rm{cm}$ signal power. Additionally, the position of sources in the MWA primary beam matters for the overall level of leakage expected in the EoR window, as the spectral behaviour of the primary beam varies dramatically across the sky. This effect can be seen in \citet{EoR-limits_Trott2020}, which demonstrated the chromatic effects of the MWA primary beam as a function of angular position by calculating the beam spectral index across a $30.72\,\rm{MHz}$ observing band. Figures $27$ to $29$ from \citet{EoR-limits_Trott2020} demonstrate the steep changes at the edges of sidelobes which have spectral indices that range from $-30$ to $30$. These Figures only capture the first order changes in the beam as a function of frequency. From \citet{Cook2021} Figure $6$ we see that for a fixed angular position the MWA primary beam can have complex polynomial like structure, not easily described by a simple power law. This spectral structure far from the main lobe of the primary beam is imparted onto radio sources, varying their spectra more rapidly with frequency. This changing structure of the MWA primary beam with position and frequency is primarily responsible for the leakage seen in the EoR window in this work.

We can assess the level of spectral leakage from CenA into the EoR window for the sidelobe observation, by comparing the expected DC power level of CenA to the power level measured in the EoR window. The apparent brightness of CenA for the sidelobe observations is $\sim10.2\,\textrm{Jy}$, which leads to an expected DC mode power of $2.54\times10^{13}$\cosmounits, after applying the appropriate conversions. The power at $k_\perp=0.01\,h\rm{\,Mpc^{-1}}$, $k_{||}=0.1\,h\rm{\,Mpc^{-1}}$ is $6.93\times10^5$\cosmounits, which is a level of leakage on the order of $0.01\%$. The apparent flux density of the SNR only sky-models for the sidelobe and null observations is comparable to the CenA apparent flux density. However, there is an order of magnitude less leakage. Performing the same calculation for the sidelobe observation with only SNRs we find a power level at $k_\perp=0.01\,h\rm{\,Mpc^{-1}}$, $k_{||}=0.1\,h\rm{\,Mpc^{-1}}$ of $7.2\times10^4$\cosmounits, for approximately $0.005\%$ leakage. Modelling and removing these sources will yield improvements by reducing leakage. This has implications for MWA EoR observations at certain pointings (not just the EoR\,2 field). In particular the EoR\,1 highband field observation from \citet{EoR-limits_Trott2020} in Figure 14, clearly has sidelobes intersecting the Galactic plane. However, this part of the Galactic Plane is not as dominated by SNRs as the part visible in the EoR2 field observations.

One important consideration is determining what the expected leakage might be for SKA-LOW observations. The individual SKA-LOW stations will have have pseudo random distributed antennas to reduces the average sidelobe gain for all the station tiles \citep{SKA}. However, as a result of the pseudo random antenna distribution, the station primary beam has two distinct regions outside the main lobe. One region with regular sidelobes close to the main lobe called the coherent region, and another region $\sim0.3\sqrt{N}$ ($N$ is the number of antennas per station) sidelobes away from the main lobe with randomly distributed sidelobes, this is called the incoherent region \citet{SKA-EoR2}. Assuming we have a similar observation of the EoR\,2 field with the future SKA-LOW array, due to the smaller field of view, CenA and the Galactic SNRs find themselves in the incoherent part of the SKA-LOW primary beam ($>30\deg$ from the main lobe). The incoherent part of the SKA-LOW primary beam has an expected power proportional to $\sim1/N=0.004$. This is confirmed for the average SKA-LOW station beam through OSKAR \citep{OSKAR} simulations of the SKA-LOW primary beam at $180$\,MHz (assuming an analytic log-dipole antenna model with no mutual coupling). The expected beam power in the incoherent region of the OSKAR simulated average primary beam was found to be $0.003$. This is coincidentally approximately the same beam power as the MWA sidelobe CenA occupies in the sidelobe observation. If we assume similar beam spectral behaviour, we would find a similar level of leakage in the EoR window for future SKA-LOW EoR\,2 field observations. Analysing how the SKA-LOW station beam changes with frequency is outside the scope of this work, however the chromatic nature of the station tiles, and the bright extended nature of radio sources in the incoherent region, will require consideration in future SKA-LOW EoR observations.

\subsection{Future Work}\label{sec:future-work}
%
%
In the process of investigating and fitting SNRs using the GLEAM cutout images, we noticed there are numerous HII regions which are bright at MWA radio frequencies. These regions also have similar sizes and scales to SNRs, and thus to the 21\,$\rm{cm}$ ionisation bubbles. Similarly to \citet{Green-SNRcat} there is a comprehensive Galactic HII catalogue containing $1442$ HII regions \citep{HII-cat_2003}. This catalogue provides diameters, and flux densities at $2.7\,\textrm{GHz}$. HII regions are relatively opaque at the lower frequencies which the MWA observes for the EoR fields. However, there are still HII regions which are bright enough to be detected at MWA frequencies and have been observed by GLEAM \citep{GLEAM}. A similar method can be applied to model the HII regions using the catalogue information as a prior.

%
%
The 1D power spectrum of the CenA only null observation in Figure \ref{fig:1D-pspec-null}, demonstrates a potential observation strategy for the EoR\,2 field, where CenA is strategically placed in a null. \citet{Sun_null} developed a method for determining the best MWA primary beam projection to place the sun in a null for a particular pointing. This could be a useful observing strategy for the EoR\,2 field going forward. This however will not be effective for Galactic Plane SNRs, since the Galactic Plane SNRs span the entire breadth of the sky. 

\subsubsection{Morphological Models}
%
%
%
%
The morphological models presented in this work are a good first step to removing their contribution from the visibilities of EoR observations, particularly for the EoR\,2 field. The CenA and Galactic SNR models have a $\sim1.5\,\rm{arcmin}$ angular resolution, which corresponds to $k_\perp=2.4\,h\rm{\,Mpc^{-1}}$. We perform a $300\,\lambda$ cutoff effectively smoothing over angular scales smaller than $\sim11.5\,\rm{arcmin}$. However, accurate models of these smaller scale components are still important. Errors on the order of a few percent for smaller scale components will be averaged over larger angular scales, introducing leakage into $k$-modes less than $0.3\,h\rm{\,Mpc^{-1}}$.

Improvements to the morphological model fitting on all relevant angular scales can be made, especially for the largest and brightest sources. Other basis functions for fitting the morphological structure besides Gaussians exist, such as shapelets \citep{shapelets_2003} which are an orthonormal set of functions based on Hermite polynomials. \citet{Woden2020} compared morphological Gaussian component models and shapelet models of the extended complex radio galaxy Fornax A. Shapelets performed better at modelling the complex smaller scale angular structure ($\theta < 11.5\,\rm{arcmin}$) of Fornax A, and could prove useful in modelling the complex structure of SNRs, as well as the intermediate scales of CenA.

\subsubsection{Centaurus A Model}

There are some important caveats regarding the morphological model of CenA, in particular the larger scale components of the outer lobes. Referring to Figure 1 of \citet{McKinley2021}, the outer lobes of CenA contain complex structure from arcminute to degree size scales. Due to the large extent of the image, the larger scales were down sampled by a factor of $19$, conserving the flux summation. This effectively removed angular structures on scales of less than $5-10\,\textrm{arcminutes}$. This reduces the complexity of the model at the cost of accuracy. As a result our model of CenA under predicts the flux density of CenA. For the model of CenA presented in this work to be useful for further EoR science the intermediate angular scales will need to be modelled appropriately.

\subsubsection{The OSIRIS Pipeline}\label{sec:future-pipeline}
%
%
The OSIRIS pipeline developed for this work is self consistent, and compares well to a similar pipeline \textsc{majick} \citep{MAJICK}. However there are several areas in which the OSIRIS pipeline can be improved. Currently OSIRIS accepts a sky-model cube, which is then Fourier transformed via an FFT to derive the Fourier sky-cube. Since Gaussians have analytic Fourier transforms it is possible to generate a Fourier sky-cube without performing an FFT. Analytic Fourier transforms of Gaussian component image cubes, would allow for a nominal speed boost, and would reduce FFT related errors \citep{Grid-Gauss-err}. However, the benefit of using an FFT is any sky-model can be input into OSIRIS. This could be incorporated as a future feature to OSIRIS, where a user can choose to perform an FFT or analytically determine the Fourier sky-cube.

The OSIRIS pipeline could also incorporate rotation synthesis. This would allow for more accurate simulations of snapshot observations; with better $(u,v)$ plane coverage. Additionally, several processes of the OSIRIS pipeline can be made parallel to increase simulation speed, which would be necessary if we were to upgrade OSIRIS to incorporate rotation synthesis. These upgrades may be unnecessary with MWA simulation packages such as \citep[WODEN;][]{Woden2022}. In future work we plan to incorporate WODEN simulations when generating observation model visibilities.

\section{Conclusions}

In this work we simulate all-sky images containing only extended radio sources such as CenA and Galactic SNRs. We use these models to determine the level of leakage in the EoR window for the MWA EoR2 field. We find that up to $\sim50-90\%$ of the complex extended sources need be subtracted from the visibilities in order to reduce leakage to a level of $\sim10-20\%$ of the expected 21\,$\rm{cm}$ signal; this is in addition to the compact point sources which are already subtracted. The leakage from these extended sources is primarily caused by widefield chromatic effects of the MWA primary beam far from the main lobe. Additionally, we find that although the future SKA-LOW primary beam is an improvement compared to the MWA, chromatic effects and leakage from widefield sources will still affect extended widefield sources. Extended widefield sources will likely need to be subtracted in order to perform EoR science with the SKA.

\section*{Acknowledgements}
This research was partly supported by the Australian Research Council Centre of Excellence for All Sky Astrophysics in 3 Dimensions (ASTRO 3D), through project number CE170100013. JHC is supported by a Research Training Program scholarship. CMT is supported by an ARC Future Fellowship under grant FT180100321.
The International Centre for Radio Astronomy Research (ICRAR) is a Joint Venture of Curtin University and The University of Western Australia, funded by the Western Australian State government. 
The MWA Phase II upgrade project was supported by Australian Research Council LIEF grant LE160100031 and the Dunlap Institute for Astronomy and Astrophysics at the University of Toronto.
This scientific work makes use of the Murchison Radio-astronomy Observatory, operated by CSIRO. We acknowledge the Wajarri Yamatji people as the traditional owners of the Observatory site. Support for the operation of the MWA is provided by the Australian Government (NCRIS), under a contract to Curtin University administered by Astronomy Australia Limited. We acknowledge the Pawsey Supercomputing Centre which is supported by the Western Australian and Australian Governments.

\section*{Data Availability}
The MWA Centaurus A radio image taken from \citealt{McKinley2021} is available through the Strasbourg Astronomical Data Center (CDS) via anonymous ftp to cdsarc.u-strasbg.fr (130.79.128.5) or via \url{https://cdsarc.unistra.fr/viz-bin/cat/J/other/NatAs}. The cutout GLEAM images used to model supernova remnants are publicly available through the GLEAM VO server \url{http://gleam-vo.icrar.org/gleam_postage/q/form} \citealt{GLEAMyr1}. The sky-models, the resulting visibilities and their power spectra were simulated via the pipeline Observational Supernova-remnant Instrumental Reionisation Investigative Simulator (OSIRIS), which is publicly available at \url{https://github.com/JaidenCook/OSIRIS}. Examples of how to replicate the sky-model and power spectrum output data arrays used in this work are available in the Github documentation. These simulations model Murchison Widefield Array Phase I data (MWA; \citealt{MWA}), available at \url{https://asvo.mwatelescope.org/}. The OSIRIS pipeline uses MWA observation metafits files to generate the primary beam for simulations, these can be downloaded at \url{https://asvo.mwatelescope.org/}. 



\bibliographystyle{mnras}
\bibliography{main} 




\appendix

\section{Cosmological Conversion}
%
%

To meaningfully understand the cosmological significance of the EoR signal we convert the $(u,v,\eta)$ coordinates and the power to be in terms of cosmological coordinates. This cosmological conversion is described by \citet{Morales2004}:

\begin{align}\label{eq:u2k}
    k_x &= \frac{2\pi u}{D_M(z)} \:\:h\,\rm{Mpc^{-1}} \\
    k_y &= \frac{2\pi v}{D_M(z)} \:\:h\,\rm{Mpc^{-1}} \\
    k_{||} &= \frac{2\pi H_0 f_{21} E(z) \eta}{c(1+z)^2} \:\:h\,\rm{Mpc^{-1}}
\end{align}

$H_0$ is the Hubble constant, $f_{21}$ is the $21\,\rm{cm}$ frequency, $z$ is the redshift, and $E(z)$ is the cosmological function given by $E(z) = \sqrt{\Omega_M(1 +z)^3 + \Omega_k (1 +z)^2 + \Omega_\Lambda}$. $D_M(z)$ is the co-moving transverse distance, which is given by \citet{Hogg2000}:

\begin{equation}\label{eq:co-moving-dist}
    D_M(z) = D_H \int_0^{z^\prime} \frac{dz^\prime}{E(z^\prime)}
\end{equation}

This is the co-moving distance and has units of $h^{-1}\,\rm{Mpc}$. This transforms our signal into cosmological units.

\subsection{Conversion Factor}

We can describe $S_\eta$ in terms of the temperature brightness using Rayleigh-Jeans law:

\begin{equation}\label{eq:Temp-brightness}
    S_\eta = \Omega\Delta\nu_f \frac{2k_b}{\lambda^2_o} T_b \:\: \rm{Jy\:Hz},
\end{equation}
$T_b$ is the temperature brightness, $\Delta\nu_f$ is the channel width in Hz, $\Omega$ is the field of in steradians. We square Equation \ref{eq:Temp-brightness}, and then normalise by the volume $\Omega\Delta\nu$, where $\Delta\nu$ is the observation bandwidth. We can relate $\Omega\Delta\nu = \theta_x \theta_y \Delta\nu$, where $\theta_x$ and $\theta_y$ are both defined in \citet{Morales2004}. \citet{Morales2004} provides a conversion for $\theta_x$ and $\theta_y$ in terms of cosmological parameters:

\begin{equation}\label{eq:co-moving-volume}
    \Omega\Delta\nu = \frac{r_x r_y \Delta r_z}{D^2_M(z)D_H} \frac{\nu_{21}E(z)}{(1+z)^2} \:\: \rm{sr\,Hz}.
\end{equation}

Note that $r_x r_y \Delta r_z = \Delta V_C$ our co-moving volume element. It can then be shown that:

\begin{equation}\label{eq:Temp-brightness-5}
    \frac{\lambda^4_o}{4k^2_b} \frac{S^2_\eta}{\Omega\Delta\nu} = \frac{\Delta\nu^2_f}{\Delta\nu^2} \frac{\Delta V_C}{D^2_M(z)D_H} \frac{\nu_{21}E(z)}{(1+z)^2} T^2_b \:\: \rm{K^2\,sr\,Hz},
\end{equation}
Rearranging we obtain our final expression:

\begin{equation}\label{eq:Temp-brightness-6}
    N^2_c (1+z)^2\frac{D^2_M(z)D_H}{\nu_{21}E(z)} \frac{\lambda^4_o}{4k^2_b} \frac{S^2_\eta}{\Omega\Delta\nu} = \Delta V_C T^2_b \:\: \rm{K^2\,Mpc^3}.
\end{equation}

From Equation \ref{eq:Temp-brightness-6} we can define the cosmological unit conversion factor from $\rm{Jy^2\,Hz^2}$ to $\rm{Mpc^3\:mK^2\:Jy^{-2}\:Hz^{-2}}$:

\begin{equation}\label{eq:conversion-factor}
    C = (1+z)^2\frac{D^2_M(z)D_H}{\nu_{21}E(z)} \frac{\lambda^4_o}{4k^2_b} \frac{N^2_c}{\Omega\Delta\nu}\times10^6 \:\:\rm{Mpc^3\:mK^2\:Jy^{-2}\:Hz^{-2}}
\end{equation}
    
\section{Thermal Noise}\label{sec:thermal-noise}

The radiometer equation for a single baselines is given by \citet{Radio-synthesis}:

\begin{equation}\label{eq:radiometer}
    \sigma = 2\frac{k_b}{A_\textrm{eff}} \frac{T_\textrm{sys}(\nu)}{\sqrt{\Delta\nu \Delta t}},
\end{equation}
$k_b = 1380.648\,\rm{Jy\,K^{-1}\,m^2}$ is Boltzmann's constant, $A_\textrm{eff} = 21.5\,\rm{m^2}$ is the effective area of the MWA tile, $T_\textrm{sys}(\nu)$ is the system temperature:

\begin{equation}
    T_\textrm{sys}(\nu) = 50 + 228\Big( \nu/150\,\textrm{MHz} \Big)^{-2.53} \,\rm{K}   
\end{equation}

\section{2D Gaussian Projection Approximation}\label{sec:app-gauss-proj}

For orthographic projections of the celestial sphere circular Gaussians will be compressed as a function of their Altitude/Zenith angle. This can be generalised in the case of an elliptical Gaussian where we have an exaggerated representation of the problem in Figure \ref{fig:lm-plane-ellipse}. In the case of Figure \ref{fig:lm-plane-ellipse} the coordinate system is the $(l,m)$ plane. The red ellipse will have some semi-major and semi-minor axis sizes $(a,b)$, a centre positioned at $(l_0,m_0)$, an azimuth angle $\phi_0$ relative to the $m$-axis, and a position angle $\theta_\textrm{pa}$ relative to the non-rotated reference frame of the ellipse.

Compression of the ellipse happens only along the radial direction, for convenience we work in the rotated reference frame which aligns with the radial direction $(l^\prime,m^\prime)$, which is rotated with respect to the azimuth angle $\theta_0$. In this case our ellipse is rotated with respect to the $m^\prime$ axis by the position angle $\theta_\textrm{pa}$. The non-rotated reference frame of the ellipse is denoted by $(l^{\prime\prime},m^{\prime\prime})$. An example of this can be seen in Figure \ref{fig:ellipse}.

\begin{figure}
    \centering
    
    \begin{tikzpicture}[scale=0.5]
        
    \begin{scope}[remember picture]
    \coordinate (O) at (0,0);
    \coordinate (eO) at (0,3.5);
    \coordinate (e1) at (-2.2,2.2);
    
    \pic["$\phi_0$", draw=black, <->, angle eccentricity=1.5, angle radius=1cm]
    {angle = eO--O--e1};
    
    \draw[line width = 0.5mm] (O) circle (4.5cm);
    
    \draw[<->] (-5, 0) -- (5, 0) node[right] {$l$};
    \draw[<->] (0, -5) -- (0, 5) node[right] {$m$};
    \draw[->] (O) -- (e1);
    
    \draw[rotate around={340:(e1)},red,line width = 0.5mm] (e1) ellipse (1cm and 0.44cm);

    \end{scope}
    \end{tikzpicture}
    \caption{(l,m) plane of the visible celestial sphere. An ellipse in red offset from the centre is located at an azimuth angle of $\phi_0$.}
    \label{fig:lm-plane-ellipse}
\end{figure}
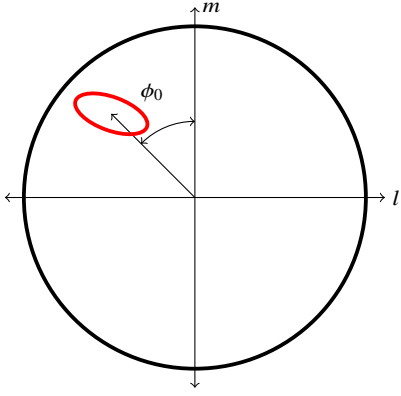

\begin{figure}
    \centering
    
    \begin{tikzpicture}[scale=0.65]
    \begin{scope}[remember picture,rotate=315]
    \coordinate (O) at (0,0);
    \coordinate (A) at (-4.5,0);
    \coordinate (B) at (0,-2);
    \coordinate (C) at (0,3.182);
    
    \draw[line width = 0.5mm] (O) ellipse (4.5cm and 2.0cm);
    \draw[->] (5, 0) -- (-5, 0) coordinate[right] (m);
    \draw[->] (0, 5) -- (0, -5) coordinate[above] (l);

    \path (O) -- (A) coordinate[pos=.02] (a1) coordinate[pos=.98] (a2);
    \draw[decorate, decoration = {brace, amplitude = 15pt, mirror, raise =4pt}, yshift = 0pt]
        (a2) -- (a1) coordinate[midway] (tl1);
    \coordinate (l1) at ([yshift=1cm]tl1);
    \path (O) -- (B) coordinate[pos=.05] (b1) coordinate[pos=.95] (b2);
    \draw[decorate, decoration = {brace, amplitude = 15pt, mirror, raise =4pt}, yshift = 0pt]
        (b2) -- (b1) coordinate[midway] (tl2);
    \coordinate (l2) at ([xshift=-.9cm]tl2);
    
    \end{scope}
    \draw[dashed, ->] (-5, 0) -- (5, 0) node[right]{$\Tilde{l}^\prime$};
    \draw[dashed, ->] (0, -5) -- (0, 5) node[above]{$\Tilde{m}^\prime \cos{\theta_0}$};
    \node[anchor=south west] at (m) {$\Tilde{m}^{\prime\prime}$};
    \node[anchor=south east] at (l) {$\Tilde{l}^{\prime\prime}$};
    \node[anchor=south west] at ([xshift=-2.cm,yshift=-2.cm]l1) {$a^\prime$};
    \node[anchor=south east] at ([yshift=-1.9cm,xshift=2cm]l2) {$b^\prime$};

    \coordinate (OO) at (0,0);
    \coordinate (AA) at (-3.182,3.182);
    \coordinate (BB) at (0,-2);
    \coordinate (CC) at (0,3.182)

    pic["$\theta_{\textrm{pa}}$", draw=black, <->, angle eccentricity=1.5, angle radius=1.25cm]
    {angle = CC--OO--AA};

    \end{tikzpicture}

    \caption{Ellipse in the non-offset rotated frame. Here the ellipse is rotated by the intrinsic position angle $\theta_{\textrm{pa}}$}
    \label{fig:ellipse}
\end{figure}
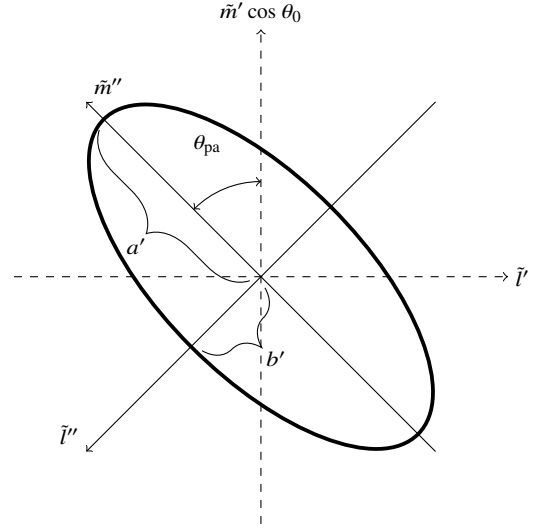
Compression of the Gaussian is a fundamentally continuous process that occurs as a function of $\cos{\theta}$, where $\theta$ is the zenith angle. Since most Gaussians in astronomy are small in angular scale we can approximate the compression, by compressing the entire $m^\prime$ axis by the value $\cos{\theta_0}$. We can then use Pythagoras theorem to determine an approximation of what the new semi-major and minor axes size will be:

\begin{align}
    a^\prime &= \sqrt{\delta l_a^2 + (\delta m_a \cos{\theta_0})^2}\\
    a^\prime &= a \sqrt{\sin^2{\theta_{\textrm{pa}}} + \cos^2{\theta_{\textrm{pa}}} \cos^2{\theta_0} }
\end{align}
\begin{align}
    b^\prime &= \sqrt{\delta l_b^2 + (\delta m_b \cos{\theta_0})^2}\\
    b^\prime &= b \sqrt{\cos^2{\theta_{\textrm{pa}}} + \sin^2{\theta_{\textrm{pa}}} \cos^2{\theta_0} }
\end{align}

Where $\delta l_a = a\sin{\theta_\textrm{pa}}$, $\delta m_a = a\cos{\theta_\textrm{pa}}$. and Where $\delta l_b = b\cos{\theta_\textrm{pa}}$, and $\delta m_b = b\sin{\theta_\textrm{pa}}$. These components are described by the uncompressed components which are derived in an uncompressed flat plane.


\bsp	
\label{lastpage}
\end{document}